\documentclass[authoryear,preprint]{elsarticle}
\usepackage{graphicx}
\usepackage{epsfig}
\usepackage{amsmath}
\usepackage{tensor}
\usepackage{natbib}
\usepackage{xcolor}
\usepackage{amsfonts}
\usepackage[normalem]{ulem}

\usepackage{colortbl}
\usepackage{mathtools}
\usepackage{euscript}
\usepackage{mathbbol}

\newcommand{\be}{\begin{equation}}
\newcommand{\ee}{\end{equation}}
\newcommand{\bea}{\begin{eqnarray}}
\newcommand{\eea}{\end{eqnarray}}

\newcommand{\nd}{\noindent}
\newcommand{\at}[2][]{#1\Bigg|_{#2}}

\begin{document}
\title{Games of multicellularity}
\author[ped]{Kamran Kaveh\corref{em}}
\author[ped,org]{Carl Veller}
\author[ped,org]{Martin A. Nowak}
\cortext[em]{Corresponding author,\\
 E-mail address: kkavehma@gmail.com (K. Kaveh).}
\address[ped]{Program for Evolutionary Dynamics, Harvard University, Cambridge, MA 02138, USA}
\address[org]{Department of Organismic and Evolutionary Biology, Department of Mathematics, Harvard University, Cambridge, Massachusetts 02138, USA}

\begin{abstract}

Evolutionary game dynamics are often studied in the context of different population structures. Here we propose a new population structure that is inspired by simple multicellular life forms. In our model, cells reproduce but can stay together after reproduction. They reach complexes of a certain size, $n$,  before producing single cells again. The cells within a complex derive payoff from an evolutionary game by interacting with each other. The reproductive rate of cells is proportional to their payoff. We consider all two-strategy games.  We study deterministic evolutionary dynamics with mutations, and  derive exact conditions for selection to favor one strategy over another. Our main result has the same symmetry as the well-known sigma condition, which has been proven for stochastic game dynamics and weak selection. For a maximum complex size of $n=2$ our result holds for any intensity of selection. For $n\geq 3$ it holds for weak selection. As specific examples we study the prisoner's dilemma and hawk-dove games. Our model advances theoretical work on multicellularity by allowing for frequency-dependent interactions within groups.

\end{abstract}
\maketitle \vspace{10pt}

\section{Introduction}

The emergence of multicellular life forms is an important step in the evolutionary history of life on earth \citep{grosberg2007evolution,bell1997size, knoll2011multiple,bonner1998origins, bonner2009first, rokas2008origins, carroll2001chance, john2009social,rainey2007unity,michod1997cooperation,michod1996cooperation,michod2001cooperation,hanschen2015evolutionary}. Multicellularity arose numerous times in prokaryotes, including in cyanobacteria, actinomycetes, and myxobacteria  \citep{grosberg2007evolution, bell1997size,schirrmeister2011origin}. Complex multicellular organisms evolved in six eukaryotic groups: animals, plants, fungi as well as brown, green and red algae. 

A comparison between simple multicellular and their relative unicellular organisms indicates multiple evolutionary transitions. These include increase in genetic complexity, cell differentiation, cell adhesion and cell-to-cell communication \citep{rokas2008origins}. Division of labor, efficient dispersal, improved metabolic efficiency, and limiting interaction with non-cooperative individuals have been suggested as advantageous traits offered by multicellularity \citep{michod2001cooperation,michod2007evolution,bonner1998origins,pfeiffer2001cooperation,pfeiffer2003evolutionary,kirk2005twelve,mora2015development}(see also \citet{grosberg2007evolution} and references therein.) 

Multicellular organisms are usually formed by single cells whose daughter cells stay together after division  \citep{bonner1998origins,koschwanez2011sucrose,maliet2015model,rossetti2011emergent}. In contrast, multicellular organisms via aggregation are formed by separate cells coming together. Staying together and coming together lead to very different evolutionary dynamics \citep{tarnita2013}, and pose different challenges for the problem of evolution of cooperation \citep{nowak2006five, nowak2010,olejarz2014evolution}. The same two modes for the evolution of complexity are also observed in the context of eusociality among insects \citep{wilson1971insect, gadagkar1994social, gadagkar2001social, hunt2007evolution}. A common route to eusociality is daughters staying with their mothers \citep{nowak2010eusociality}, but there is also the coming together of different individuals in the formation of new colonies \citep{wilson1971insect, gadagkar2001social}.

Here, we carry out a theoretical study of the dynamics underlying the evolution of multicellularity. Previous studies of such dynamics, both theoretical and experimental, have often been carried out under the assumption that within-group fitnesses derive from a simple, additive cooperative dilemma. For example, cells producing ATP from an external energy resource might do so with high yield but low rate, or with low yield but high rate \citep{pfeiffer2001cooperation}. In the context of a group of cells trying to make use of an energy resource, the former behaviour characterizes cooperators, and the latter defectors, because the benefits of a high rate of resource use accrue to the individual cell, while the costs of inefficient resource use accrue more broadly within the group \citep{pfeiffer2003evolutionary}. If the costs accrue equally to all group members, the strategic problem within the group can be conceptualized as an additive public goods game. Many other models of the evolution of multicellularity can be conceptualized in the same way \citep{penn2012can}. For example, the aggregation of biofilms in \textit{Pseudomonas} bacteria involves the production, costly to individual providers, of the components of an extracellular matrix and other substances \citep{davies1995regulation, matsukawa2004putative, diggle2006galactophilic}.

This assumption reduces the strategic conflicts within each multicellular unit to a very simple, frequency-independent form \citep{michod1999darwinian}. Because a group's reproductive success is shared equally among its constituents (no matter their type), the only within-group conflict involves the constant cost to cooperation. 

\begin{figure}
\begin{center}
\epsfig{figure=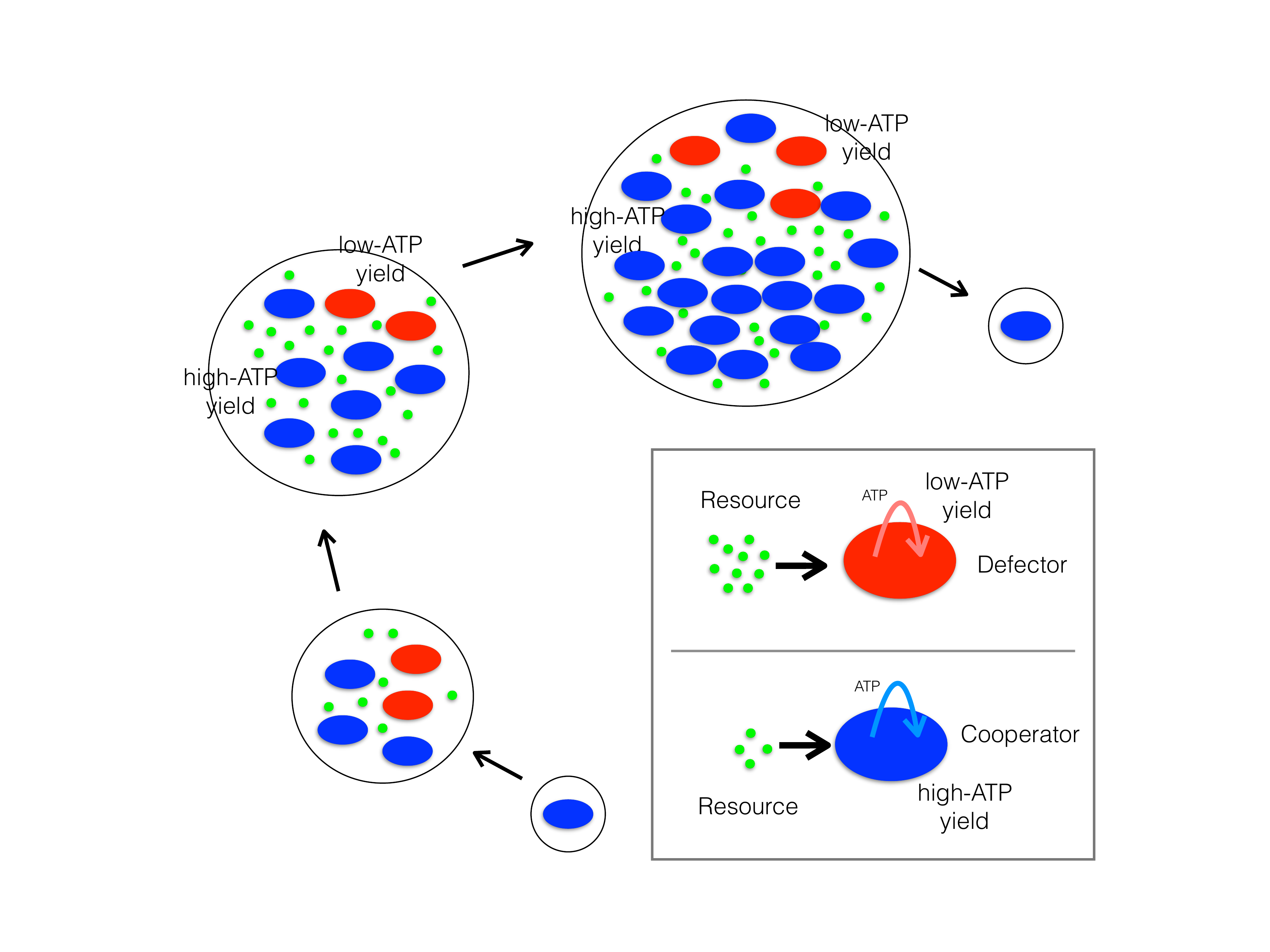, height=270pt,width=350pt,angle=0}
\end{center}
\caption{Growth of a simple multicellular complex containing two competing metabolic phenotypes \citep{pfeiffer2001cooperation}. The cooperator phenotype (blue) uses the limited food resource to produce ATP with high efficiency but low rate; the defector phenotype produces ATP with low efficiency but high rate. If the benefits of efficient resource use are shared equitably among the whole group, and the benefits of a high rate of resource use are enjoyed by individual cells, then the within-group conflict is a linear public goods game. In this case, the results of interactions within the group are frequency-independent, and defectors always grow as a proportion of the group. On the other hand, if the benefit of efficient resource use is shared more locally, then within-group strategic interactions are more complex. Now, cooperators can increase as a proportion of the group if they typically interact with cooperators, which can occur, for example, if a viscous spatial structure governs interactions (pictured).}
\label{atp}
\end{figure}

This is not realistic in many scenarios. In the example of ATP production described above, if the benefits of efficient resource use accrue more locally than to the whole group (for example, to pairs of interacting cells within the group), then the strategic interactions among cells are more complicated than a linear public goods game (Fig.\ref{atp}). Without taking this into account (i.e., assuming that the benefits produced by cooperators are shared evenly among group members), it would seem that defectors should always be at an advantage within the group. But once the strategic complexity of local interactions is taken into account, then cooperators can have a within-group advantage if most of their interactions within the group are with fellow cooperators (Fig.\ref{atp}).

Another example where strategic interaction within the group is important is when certain cell types are preferentially found in the reproductive propagules emitted by the group. Thus, in multicellular clusters of the yeast \textit{Saccharomyces cerevisiae}, experimentally selected for by gravity-based methods, some cells (cooperators) undergo apoptosis to destabilise the multicellular unit and create new propagules; having apoptosed, they cannot themselves be in these propagules \citep{ratcliff2012experimental, pentz2015apoptosis}. 

Another example involves cells that either aggressively or passively try to sequester resources for themselves; if the presence of many aggressive types involves a destructive cost to them, then the within-group conflict resembles a hawk-dove game. Because the within-group conflicts are frequency-independent in this example, their effects in the context of the evolution of multicellularity cannot be understood under a linear public goods conceptualization.

To put it concisely, the evolution of multicellularity is often studied in a framework that does not adequately account for the \emph{interactions} of cells within a group. In this paper, we place the evolution of multicellularity into an explicitly game-theoretic framework. Evolutionary game dynamics is the study of frequency dependent selection \citep{smith1982evolution, hofbauer1998evolutionary, nowak2006evolutionary}. The success of a genotype (or phenotype or strategy) depends on the frequency of different genotypes in the population. Evolutionary game dynamics was initially studied in well-mixed and infinitely large populations using deterministic differential equations \citep{hofbauer1998evolutionary,smith1982evolution,weibull1997evolutionary}. More recently it has moved to finite population sizes using stochastic dynamics \citep{nowak2006evolutionary, taylor2004evolutionary, traulsen2009stochastic}.
Evolutionary games are also studied in structured populations
\citep{nowak1992evolutionary, page2000spatial,hauert2004spatial, ohtsuki2006simple, szabo2000spatial, tarnita2009strategy, tarnita2009evolutionary,hauert2012evolutionary,langer2008spatial,antal2009evolution,allen2015games,cooney2015assortment}.  

A game-theoretic approach to the evolution of multicellularity allows us to generalize the traditional framework by accounting for frequency-dependent competition within multicellular units. 

The primary goal of our paper is to understand how the population structure of simple multicellularity affects the outcome of biological games. Previous studies have explored the evolutionary emergence of staying together \citep{tarnita2013} in the context of diffusible public goods \citep{olejarz2014evolution} and in stochastic dynamics \citep{ghang2014stochastic}. Here we study deterministic evolutionary dynamics in a population where staying together has already evolved.

In our model, single cells divide, but the two daughter cells can stay together after cell division. These cells may undergo further division until the complex reaches a specified maximum size. Thereafter, the complex does not grow further but produces single-cellular offspring, which subsequently form new complexes. Within a complex, cells interact according to a biological game. This means they derive payoffs which affect their reproductive rate. We consider natural selection acting on two types of cells (or strategies), determined by their genotype. 

We include mutation between the two types, assumed to occur during cell division. Each offspring adopts its parent's type with probability $1-u$ and changes to the other type with probability $u$. We shall be interested both in low rates of mutation (corresponding, for example, to nucleotide substitutions) and in very high rates of mutation (for example, genetic switches, epigenetic marking, or structural mutations deriving from a modular genetic architecture -- a fuller discussion of these is provided in the Discussion section). In the absence of mutation, $u = 0$, one of the two types is bound to take over the whole population (fixation). With mutation, $0 < u < 1$, the system goes to a mutation-selection equilibrium in which both types are present. We can say that selection favours one type if it is more abundant at equilibrium \citep{antal2009strategy,tarnita2009strategy,allen2014measures}.

For the simple case of a maximum complex of size $n = 2$, we derive exact solutions for the model and for the condition that a strategy (or type) is favored for any intensity of selection. Subsequently, we derive results for weak selection for any maximum complex size, $n\geq 3$. Our results have the same symmetry as the well known $\sigma$-condition for evolutionary graphs and evolutionary sets \citep{tarnita2009strategy}.
The $\sigma$-condition is an algebraic condition that describes when selection favours one strategy over another. The $\sigma$-condition holds for any population structure that treats the two strategies symmetrically for stochastic evolutionary dynamics and for weak selection. For more references on $\sigma$-conditions, see \citet{tarnita2011multiple,mcavoy2015structure,nathanson2009calculating,nowak2010,allen2012mutation}. In our case a $\sigma$-type condition arises for a deterministic evolutionary process. For $n=2$ it  holds for any intensity of selection. For $n \geq 3$ it holds for weak selection.

We apply our finding to evolution of cooperation and to the hawk-dove game. We observe that the population structure of simple multicellularity can easily favor cooperation over defection and doves over hawks.

The paper is structured as follows. In Section 2 we introduce the basic model for a maximum complex size of $n=2$ and state the main results. In Section 3 we study evolution of cooperation and the hawk-dove game. We also discuss how the 
average fitness at equilibrium depends on the mutation rate in these two games. In Section 4, we discuss the model for $n \geq 3$ and derive the recurrence relations for equilibrium solutions. In Section 5, we derive generalized $\sigma$-condition for weak selection and show that the results match numerical solutions. In Appendices A and C, we discuss technical details for the $n = 2$ and $n = 3$ analytical solutions. In Appendix B, we address evolutionary stability for $n = 2$. In Appendix D, we discuss the selection condition for an unstructured game which corresponds to $\sigma=1$ \citep{antal2009strategy}. 

\section{Model and results for maximum complex size $n=2$}

We consider a model with two types of cells,  0 and 1. Both cell types divide and reproduce. They also have a chance of staying together to form complexes. The two-cell complexes can be either 00, 01 or 11. For the moment we limit ourselves to a model with maximum complex size, $n=2$. If a cell in a complex of size two reproduces, the daughter cell leaves and joins the pool of single cells. 

During each cell division, there is a probability of mutation. An offspring of a type 0 cell mutates to a type 1 cell with probability $u$ or remains a type 0 cell with probability $1-u$. We assume symmetric mutations: the probability to mutate from 0 to 1 is the same as from 1 to 0. 

\begin{figure}
\begin{center}
\epsfig{figure=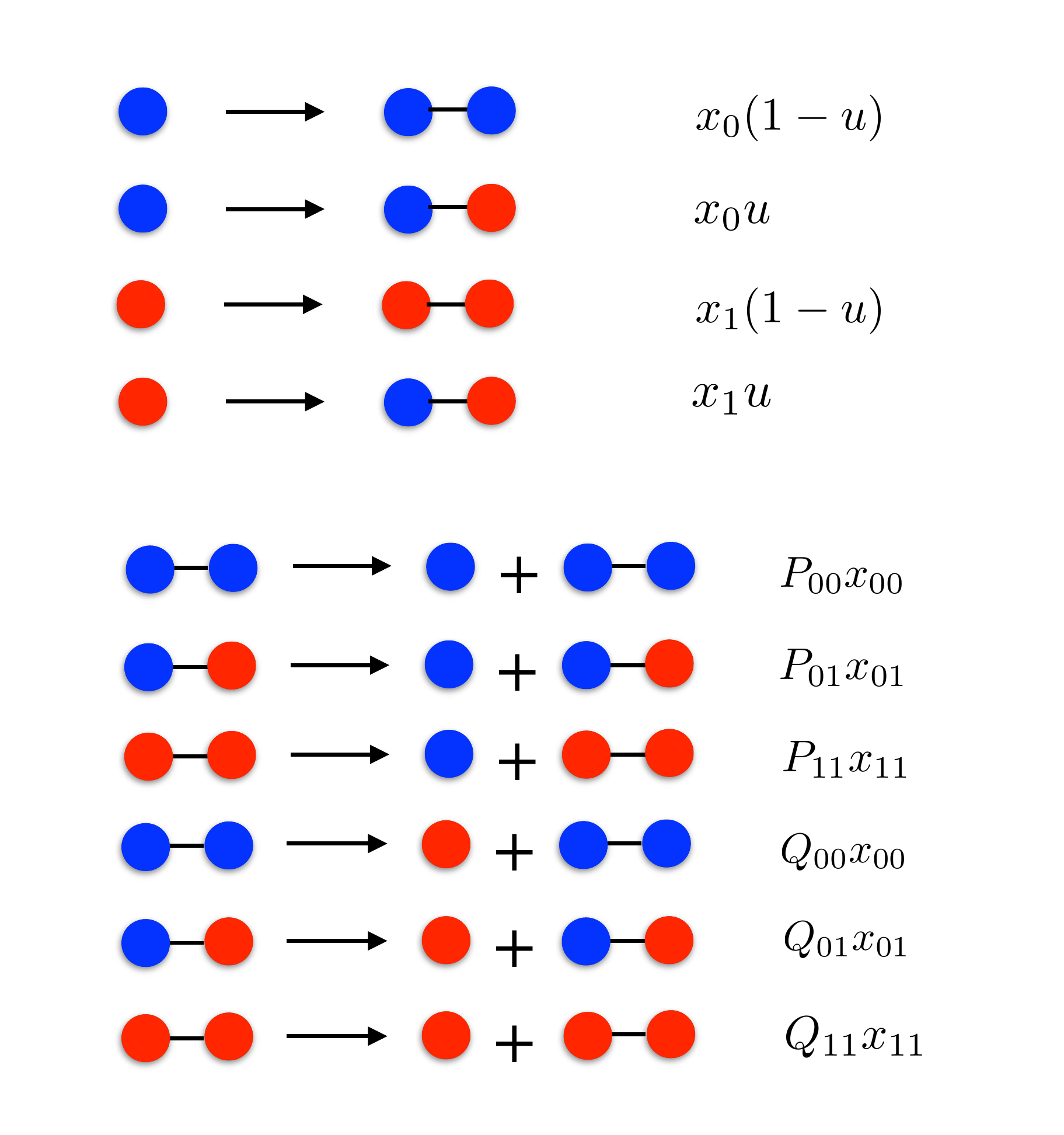, height=230pt,width=210pt,angle=0}
\end{center}
\caption{All possible events in a game of multicellularity if the maximum complex size is $n=2$. Type 0 cells
are shown in blue, and type 1 cells are in shown red. The frequencies of single cells of type 0 and 1 are denoted by $x_{0}$ and $x_{1}$. The frequencies of 00, 01 and 11 complexes are denoted by $x_{00},x_{01},x_{11}$, respectively. The division rates of single cells of both types are set to unity.
The coefficients $P_{00}, P_{01}$ and $P_{11}$ denote the rates at which these complexes generate type 0 cells. 
The coefficients $Q_{00}, Q_{01}$ and $Q_{11}$ denote the rates at which these complexes generate type 1 cells. 
}
\label{reaction1}
\end{figure}

The division rate of single cells is set to unity. The model is depicted in Fig. \ref{reaction1}. Denoting the abundances of type 0 and type 1 cells by $x_{0}$ and $x_{1}$ and denoting the abundances of the complexes 00, 01 and 11 by $x_{00},x_{01}$ and $x_{11}$, we can write the  dynamics of these five populations as

\bea
\dot{x}_{0} &=& P_{00}x_{00} + P_{01}x_{01}+P_{11}x_{11} -x_{0} - x_{0}\phi\nonumber\\
\dot{x}_{1} &=& Q_{00}x_{00}+Q_{01}x_{01}+Q_{11}x_{11}-x_{1}-x_{1}\phi\nonumber\\
\dot{x}_{00}&=& (1-u)x_{0} - x_{00}\phi\nonumber\\
\dot{x}_{01} &=& u(x_{0}+x_{1}) - x_{01}\phi\nonumber\\
\dot{x}_{11}&=& (1-u)x_{1}-x_{11}\phi
\label{gm-mult}
\eea

The coefficients $P_{00}, P_{01}$ and $P_{11}$ denote the rates at which these complexes generate type 0 cells, while 
the coefficients $Q_{00}, Q_{01}$ and $Q_{11}$ denote the rates at which these complexes generate type 1 cells. 
These coefficients  depend on the payoff derived from the game, the intensity of selection and the mutation rate. They are as follows 

\bea
P_{00} &=& 2(1 + wa)(1-u)\nonumber\\
P_{01} &=& (1+ wb)(1-u) + (1+ wc) u \nonumber\\
P_{11} &=& 2(1 + wd)u\nonumber\\
Q_{00} &=& 2(1+wa)u\nonumber\\ 
Q_{01} &=& (1+wb)u + (1+wc)(1-u)\nonumber\\
Q_{11} &=& 2(1+wd)(1-u)
\label{coeff}
\eea

The parameters $a,b,c,d$ are the elements of the $2\times 2$ payoff matrix

\bea
\bordermatrix{~ & {\rm 0} & {\rm 1} \cr
                  {\rm 0} & \displaystyle a & b\cr
                  {\rm 1} & c & \displaystyle d}
\eea

In each complex a type 0 cell obtains payoff $a$ from another type 0 cell, and $b$ from a type 1 cell. Similarly, a type 1 cell obtains payoff $c$ from a type 0 cell, and $d$ from a type 1 cell. The game interaction occurs only between cells within the same complex. The intensity of selection is denoted by $w$ and measures how much the payoff of the game contributes to the fitness. 

Note that the reproductive rate of a cell, which multiplies the mutation rate, must always be non-negative. Therefore we require $1+w \min\{a,b,c,d\} \geq 0$. If some entries of the payoff matrix are negative then these conditions limit the maximum intensity of selection.

The average fitness, $\phi$, is obtained from the constraint that the relative abundances sum up to unity, $x_{0} + x_{1} + 2(x_{00} + x_{01} +x_{11})=1$. We have
\be
\phi = (P_{00}+Q_{00})x_{00} + (P_{01}+Q_{01})x_{01}+(P_{11}+Q_{11})x_{11} + x_{0}+x_{1}
\label{phi}
\ee

The equilibrium abundances are obtained by setting all time derivates in Eq. \ref{gm-mult} equal to zero. The solutions can be expressed in terms of ratio of type 1 to type 0 singlets $\eta^{\star}=x^{\star}_{1}/x^{\star}_{0}$ and the value of average fitness at equilibrium, $\phi^{\star}$:

\bea
x^{\star}_{0} &=& \frac{\phi^{\star}}{2 + \phi^{\star}}\frac{1}{1+\eta^{\star}}\nonumber\\
x^{\star}_{1}&=& \frac{\phi^{\star}}{2 + \phi^{\star}}\frac{\eta^{\star}}{\eta^{\star} + 1}\nonumber\\
x^{\star}_{00}&=&\frac{1-u}{2+\phi^{\star}}\frac{1}{1+\eta^{\star}}\nonumber\\
x^{\star}_{01}&=&\frac{u}{2 +\phi^{\star}}\nonumber\\
x^{\star}_{11}&=&\frac{1-u}{2+\phi^{\star}}\frac{\eta^{\star}}{1+\eta^{\star}}\nonumber\\
\label{equil}
\eea

\nd Eq. \ref{equil} and the values of $\phi^{\star}$ and $\eta^{\star}$ in terms of game payoffs and mutation rate are derived in Appendix A (Eqs. \ref{ABC}-\ref{equil-app}). The total equilibrium abundances  of type 0 and type 1 cells are

\bea
x^{\star}_{\rm tot,0}&\equiv& x^{\star}_{0} + 2x^{\star}_{00} + x^{\star}_{01}\nonumber\\
&=&\frac{1}{1+\eta^{\star}}\frac{1}{2+\phi^{\star}}\big( \phi^{\star}+2(1-u)+u(1+\eta^{\star})\big)\nonumber\\
x^{\star}_{\rm tot,1}&\equiv& x^{\star}_{1} + 2x^{\star}_{11} + x^{\star}_{01}\nonumber\\
&=&\frac{1}{1+\eta^{\star}}\frac{1}{2+\phi^{\star}}\big( \phi^{\star}\eta^{\star}+2\eta^{\star}(1-u)+u(1+\eta^{\star})\big)
\label{xtot0-xtot1}
\eea

\nd A strategy is favored by selection, if its equilibrium frequency is greater than what it is in the neutral case. Here the neutral abundance of type 0 and type 1 is 1/2. Thus, the condition for type 0 to be selected over type 1 is that the total number of type 0 cells is larger than total number of type 1 cells at equilibrium

\be
x^{\star}_{\rm tot,0} > x^{\star}_{\rm tot,1}
\label{xtot-01}
\ee

\nd Substituting from Eq. \ref{xtot0-xtot1}, and using the fact that the average fitness is always positive, $\phi^{\star} > 0$, we arrive at the condition

\be
\eta^{\star} < 1
\label{eta-condition}
\ee

\nd Substituting for $\eta^{\star}$ from Eq. \ref{phi-eta1}, and Eq. \ref{ABC}, Eq. \ref{eta-condition} becomes

\be
\Big(u-\frac{1}{2}\Big)\Big( \frac{1-u}{u}a+b - c-\frac{1-u}{u}d\Big) <  0
\label{sigma}
\ee

\nd There are two zeros for the equality. We denote them $u_{1}$ and $u_{2}$. We have $u_{2}=1/2$ and

\be
u_{1} =\frac{a-d}{a-d+c-b}
\label{uc}
\ee

\nd If $u$ is outside the interval of $(u_{1},u_{2})$ (or $(u_{2},u_{1})$ if $u_{2} < u_{1}$) then type 0 is favored. Inside this interval, however,
type 1 is selected. 

For $u < u_{1}$ (assuming $u_{1} < u_{2}$) the condition for type 0 to be favored simplifies to

\be
\sigma a + b > c + \sigma d
\label{sigma0}
\ee

\nd Here $\sigma = (1-u)/u$ is only a function of $u$ and independent of the payoff values. This condition, also known as the $\sigma$-condition, has been discussed in the past for other population structures for stochastic dynamics \citep{tarnita2009strategy, tarnita2011multiple}. The value $\sigma=1$ leads to the risk dominance condition in unstructured evolutionary games; see Appendix D, as well as \citet{harsanyi1988general, antal2009strategy}. Notice that our result for multicellular games holds for any selection intensity. 

The same result can be intuitively argued in the weak selection limit. Inside a complex, fitness gains $\delta f_{0}$ and  $\delta f_{1}$ of type 0 and type 1 cells  are

\bea
\delta f_{0} &=& 2a x^{\star}_{00} +  bx^{\star}_{01}\nonumber\\
\delta f_{1} &=& 2d x^{\star}_{11} + cc^{\star}_{01}
\label{weak-w}
\eea

\nd The condition for type 0 strategy to be selected is

\be
\delta f_{0} > \delta f_{1}
\label{df}
\ee

\nd If the finesses of the two phenotypes (0 and 1) were the same we would have equal abundances for type 0 and type 1 cells in the system. At weak selection we can replace $x^{\star}_{00}, x^{\star}_{01}, x^{\star}_{11}$ with $w=0$ abundances $\hat{x}^{\star}_{00} = \hat{x}^{\star}_{11}= (1-u)/6, \hat{x}^{\star}_{01}=u/3$. Here $\hat{x}$ denotes the $w\to 0$ limit. Substituting these values into Eq. \ref{df} we recover the $\sigma$-condition, Eq. \ref{sigma0}. The interplay between payoffs and mutation rates can be readily seen from Eqs.\ref{weak-w} and \ref{df}, where values of frequencies at zero selection intensity are determined solely by $u$ whereas the fitness gains per cell are determined by the payoffs. To see the connection between the two derivations, we can verify that Eq. \ref{df} is equivalent to

\be
P_{00}\hat{x}^{\star}_{00}+P_{01}\hat{x}^{\star}_{01}+P_{11}\hat{x}^{\star}_{11} > Q_{00}\hat{x}^{\star}_{00}+Q_{01}\hat{x}^{\star}_{01}+Q_{11}\hat{x}^{\star}_{11}
\label{sigma-PQ-n2}
\ee

\nd for $u < u_{1}$. This is basically the weak selection limit of the $x^{\star}_{0} > x^{\star}_{1}$ condition as in Eq. \ref{eta-condition}. 

In the above discussion we have assumed a positive mutation rate $u > 0$. It was implied that the $\sigma$-condition holds for some mutation rate $0<u<1$. In fact, from Eq. \ref{sigma}, the condition for dominance of type 0 for $u \to 0$ simplifies to $a > d$. As shown in Appendix B this is  the condition for type 0 to be an evolutionary stable strategy (ESS) at weak selection. If $a < d$ then the type 1 strategy is ESS, and for $u < u_{1}$ type 1 is selected. 

The uniqueness of solutions for Eqs.\ref{gm-mult}-\ref{equil} is true for $u \neq 0$. At $u=0$, there are two equilibrium solutions for the model. One of the two fixed-points consists of all type 0 cells and and the other is all type 1. Only one of the two strategies can be ESS. Thus for $u=0$ we
always have an attractive (Lyapunov stable) fixed point (type 0 if $a > d$) and a saddle-point corresponding to the other strategy.  For unstructured evolutionary games, the $\sigma$-condition simplifies to the condition for risk dominance, $a + b > c + d$, and the condition for evolutionary stability is the same as the Pareto efficiency, $a>d$.

\section{Examples}
\subsection{Cooperation}
Consider the payoff matrix for a simplified game of cooperation

\bea
\bordermatrix{~ & {\rm C} & {\rm D} \cr
                  {\rm C} & \mathcal{B}-\mathcal{C} & -\mathcal{C} \cr
                  {\rm D} & \mathcal{B} & 0 \cr}
\label{coop-payoffs}
\eea

\begin{figure}[h!]
\begin{center}
\epsfig{figure=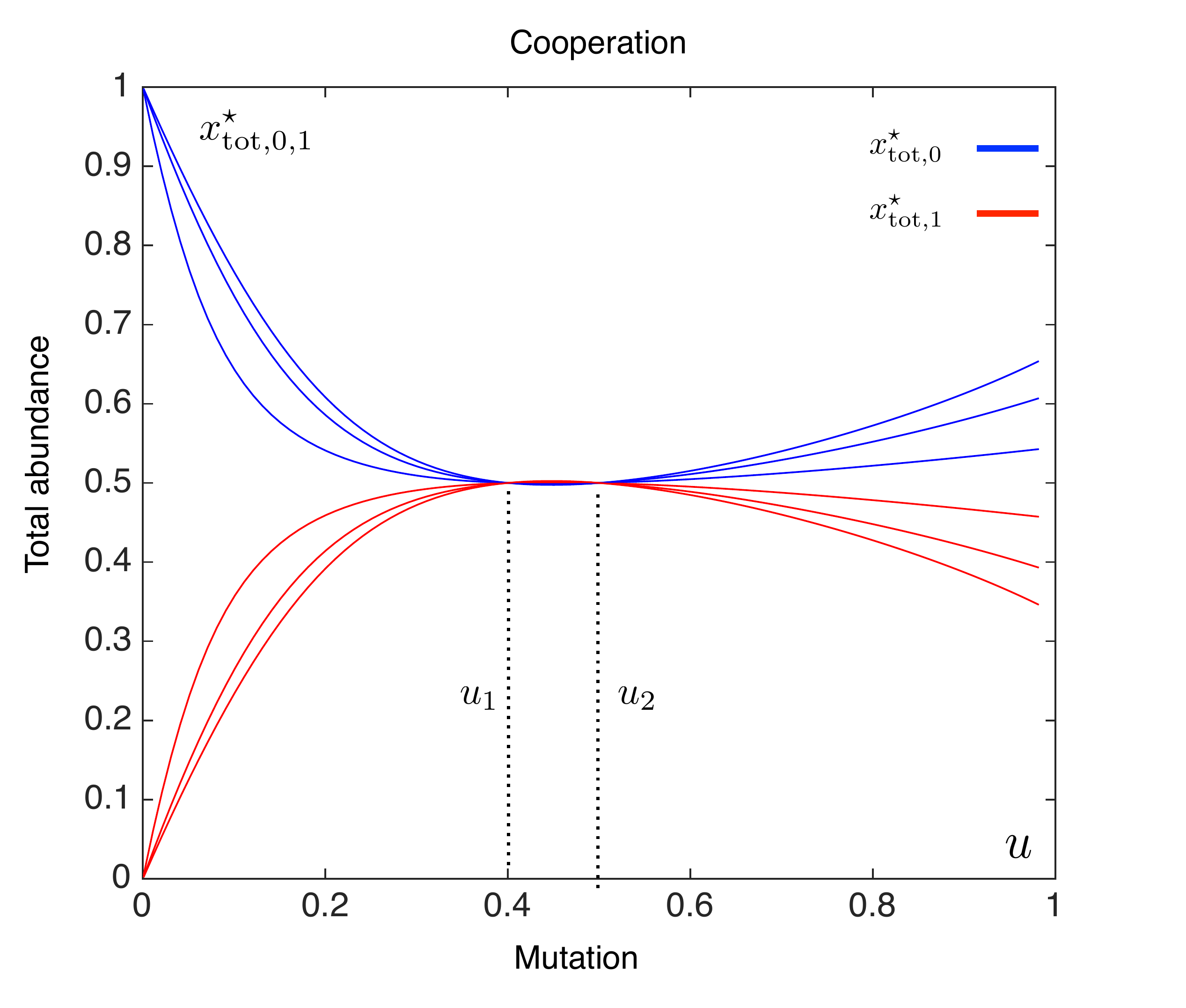, height=190pt,width=240pt,angle=0}
\epsfig{figure=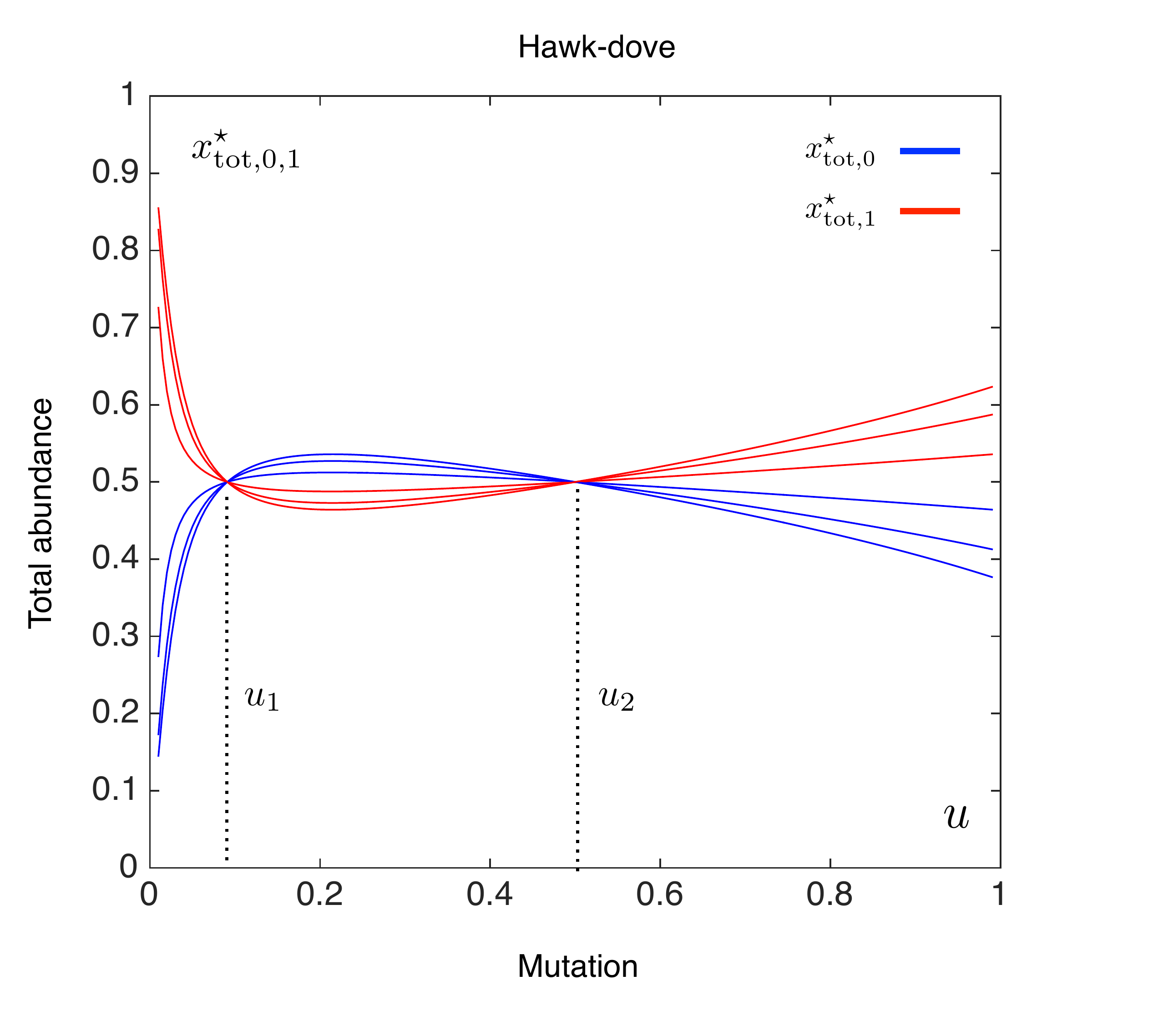, height=190pt,width=240pt,angle=0}
\end{center}
\caption{Numerical equilibrium solutions for the game of cooperation and the hawk-dove game. The total abundances  of type 0 cells (blue) and  type 1 cells (red) are plotted as a function of the mutation rate, $u$.  The parameters are $\mathcal{B} = 5$ (benefit) and $\mathcal{C}=1$ (cost). Different graphs correspond to varying selection intensity $w=0.1,0.3,0.5$; the curvature increases with $w$.}
\label{xtot_uvar}
\end{figure}

\begin{figure}[h!]
\begin{center}
\epsfig{figure=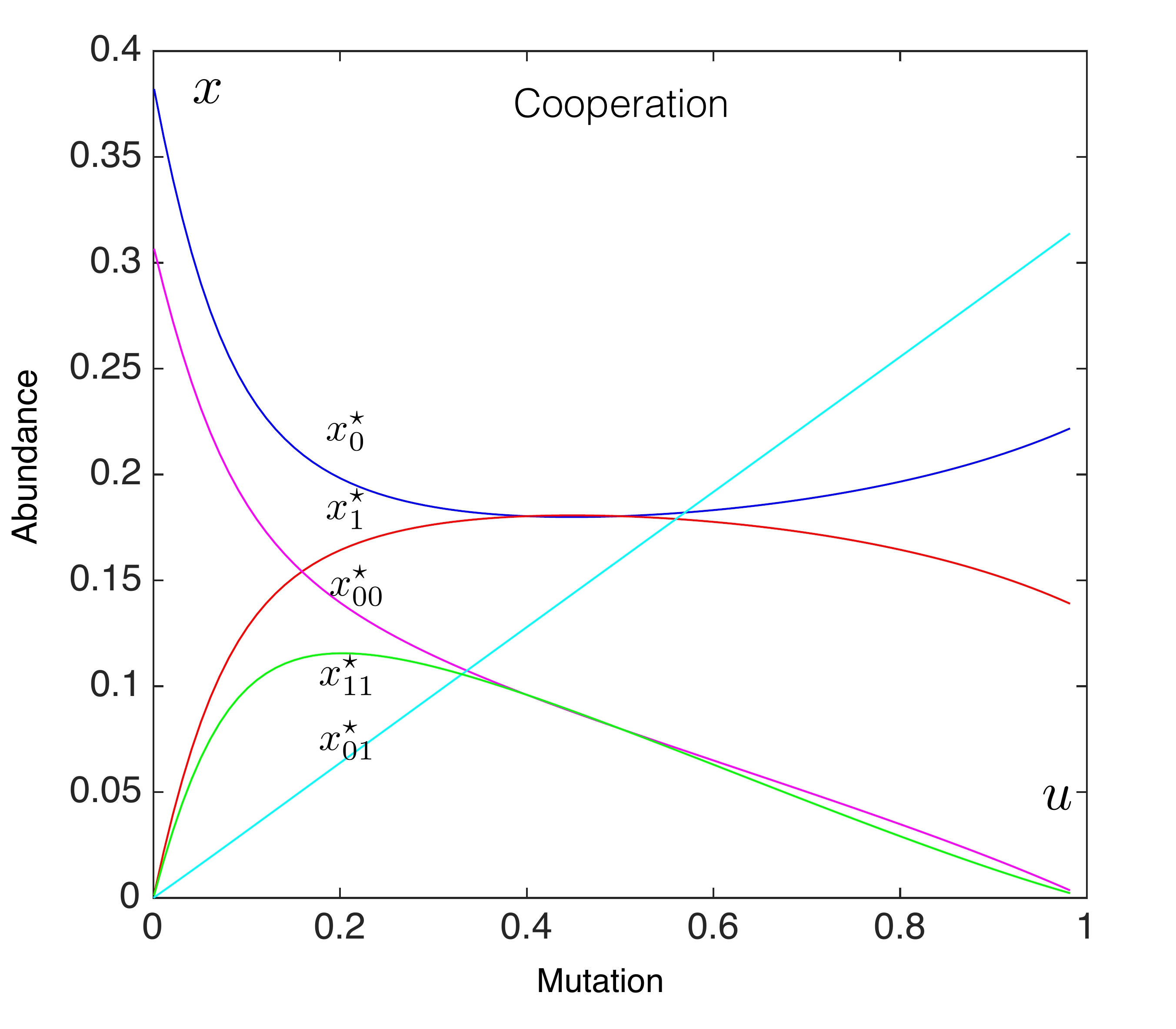, height=190pt,width=240pt,angle=0}
\epsfig{figure=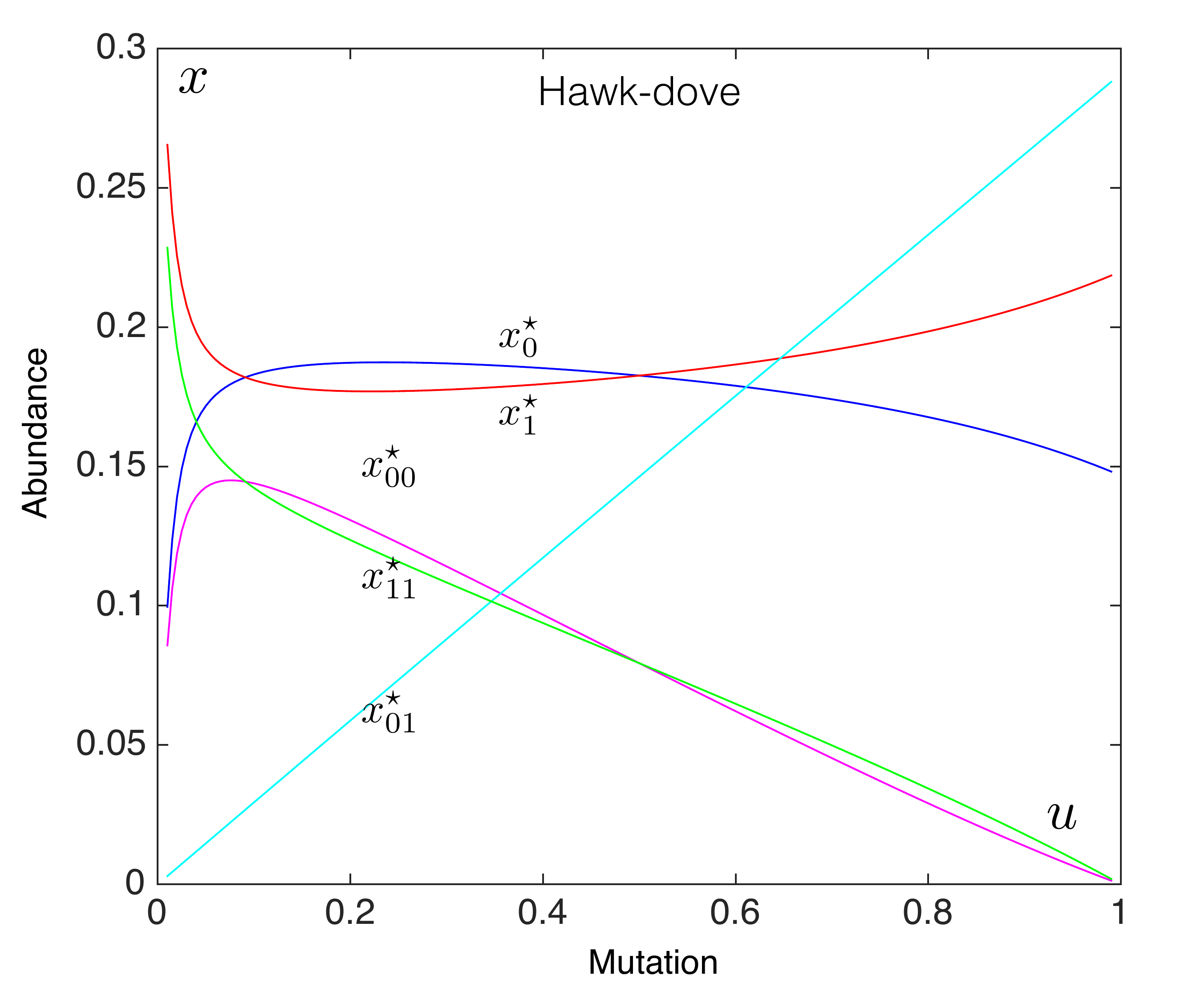, height=190pt,width=240pt,angle=0}
\end{center}
\caption{Equilibrium solutions for all singlet and doublet abundances in the game of cooperation and the hawk-dove game. All the abundances are plotted as a function of the mutation rate, $u$. Parameters are $\mathcal{B} = 5$, $\mathcal{C}=1$ and $w=0.1$.}
\label{Allx_uvar}
\end{figure}

\begin{figure}
\begin{center}
\epsfig{figure=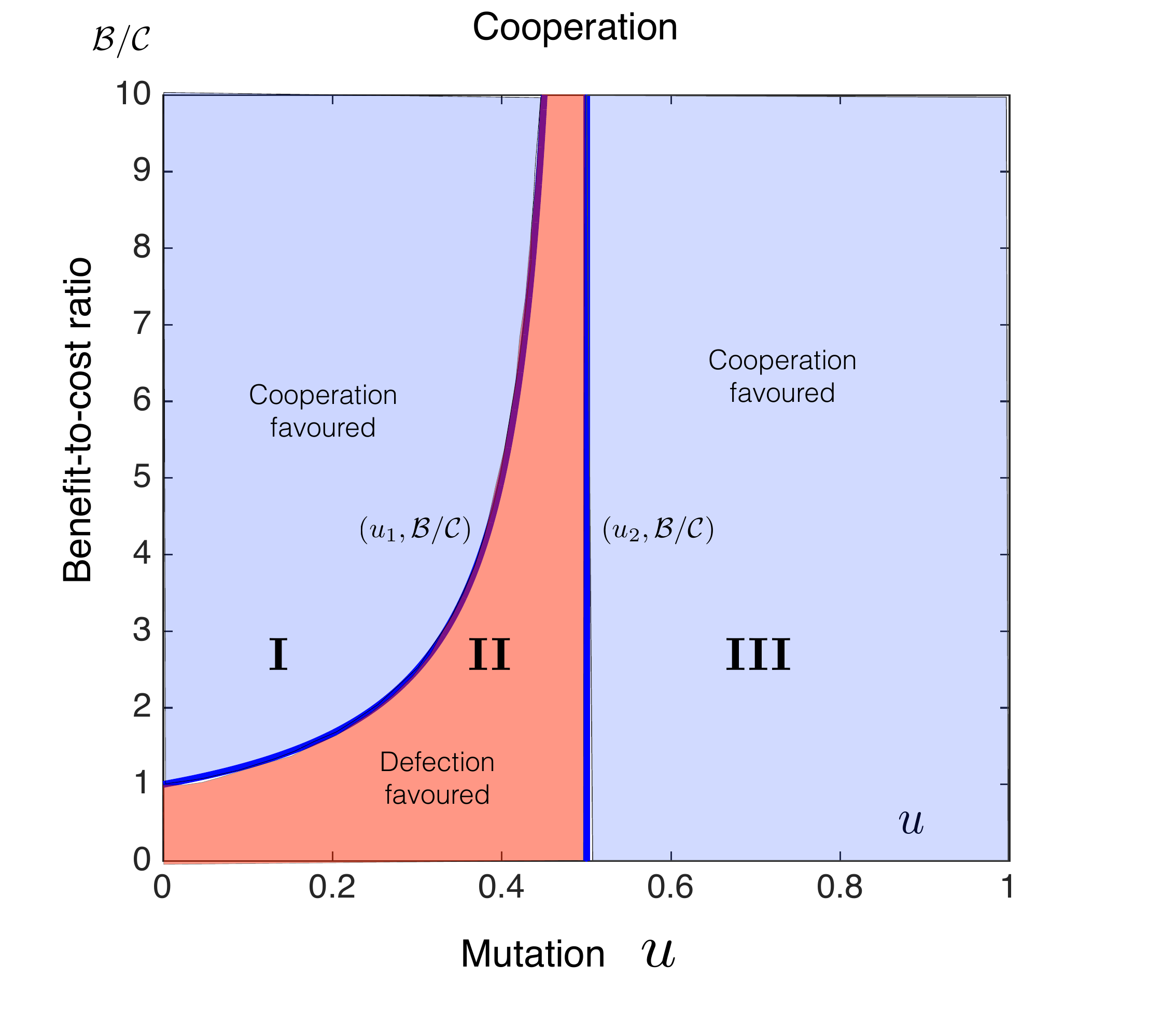, height=210pt,width=250pt,angle=0}
\epsfig{figure=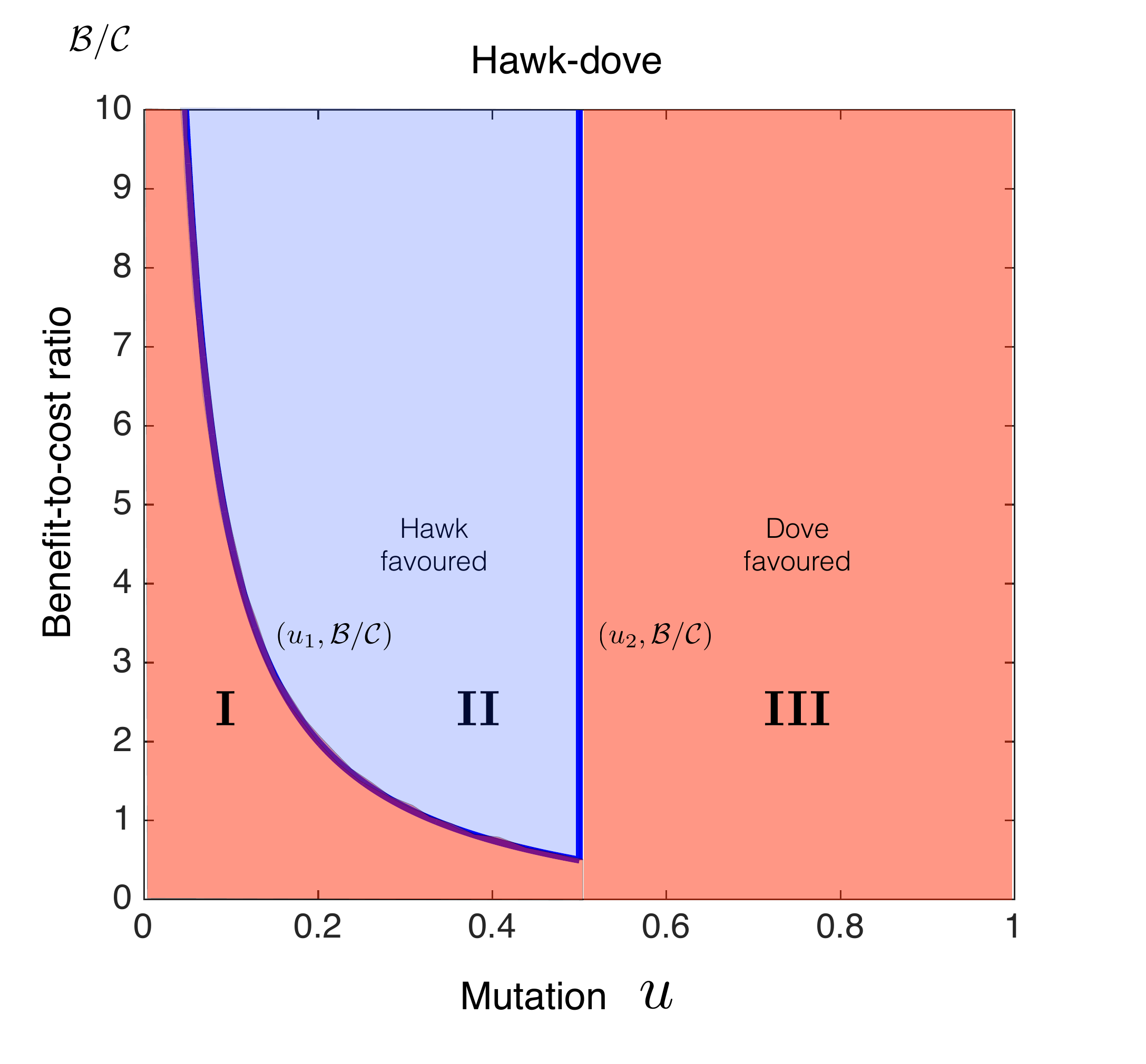, height=210pt,width=250pt,angle=0}
\end{center}
\caption{Phase diagram of the cooperation game (top) and the hawk-dove game (bottom). In the cooperation game, cooperators are favored for $u < u_{1}$ and $u_{2}=1/2 < u$ and $\mathcal{B} > \mathcal{C}$ (regions I and III). In the hawk-dove game, hawks are favored for $u_{1} < u < u_{2}=1/2$ and $\mathcal{B} > 2\mathcal{C}$ (region II).}
\label{gm-phase}
\end{figure}

\nd Here $\mathcal{B}$ and $\mathcal{C}$ indicate benefit and cost values respectively.. Type 0 cells denote the cooperator strategy, C. Type 1 cells denote the defector strategy, D. Inside a complex, a type 0 cell pays a cost, $\mathcal{C}$, and provides a benefit, $\mathcal{B}$, to the other cell. A type 1 cell pays no cost and provides no benefit.

At zero mutation and for $\mathcal{B} > \mathcal{C}$ the dynamics is driven by pure complexes, 00 and 11. Here cooperator complexes are advantaged over defector complexes. Mixed complexes, 01, are bound to become extinct. In this population structure, cooperators are evolutionarily stable. This is in contrast to evolution of cooperation in unstructured populations where defectors are stable.

For $0<u < 1/2$, the $\sigma$-condition can be written in terms of cost and benefit values as

\be
\frac{\mathcal{B}}{\mathcal{C}} > \frac{1-u}{u} 
\label{uc-coop}
\ee

\nd The critical mutation rate is

\be
u_{1} = \frac{1}{2}\Big(1 - \frac{1}{\mathcal{B}/\mathcal{C}}\Big) < \frac{1}{2}=u_{2}
\label{coop-uc}
\ee

\nd Values of $x^{\star}_{\rm tot,0}$ and $x^{\star}_{\rm tot,1}$ as functions of $u$ are plotted in Fig. \ref{xtot_uvar} for parameters $\mathcal{B}=5$ and $\mathcal{C}=1$. For these parameter values we observe $u_{1}=0.4$ and $u_{2}=0.5$, in agreement with Eq. \ref{coop-uc}. In Fig. \ref{Allx_uvar} frequencies of all populations (singlets and complexes) are plotted as a function of $u$ for the same cost and benefit values and $w=0.1$. We see that the condition $x^{\star}_{\rm tot,0} >  x^{\star}_{\rm tot,1}$ coincides with $x^{\star}_{0} > x^{\star}_{1}$ in agreement with Eq.\ref{eta-condition}. 
Eq. \ref{coop-uc} describes the phase boundary in the space of mutation and benefit-to-cost ratio. This is depicted in Fig. \ref{gm-phase}. The region between the curves $(u_{1},\mathcal{B}/\mathcal{C})$ and $(u_{2},\mathcal{B}/\mathcal{C})$ is where the defector strategy is selected. The two phase boundaries never meet as $u_{1}=u_{2}=1/2$ does not have a positive finite solution for $\mathcal{B}/\mathcal{C}$. In fact $u=u_{2}$ is the vertical asymptote of the critical benefit-to-cost ratio as a function of $u$. 

\subsection{Hawk-dove}
Now consider a hawk-dove game given by the payoff matrix

\bea
\bordermatrix{~ & {\rm H} & {\rm D} \cr
                  {\rm H} & \displaystyle \frac{\mathcal{B}-\mathcal{C}}{2} & \mathcal{B} \cr
                  {\rm D} & 0 & \displaystyle \frac{\mathcal{B}}{2}}
\eea

Type 0 is hawk (H) and type 1 is dove (D). Inside 00 complexes each hawk gains payoff $\mathcal{B}/2$ and pays the cost $\mathcal{C}/2$. In a 01 complex, a hawk gains payoff $\mathcal{B}$ while a dove does not gain from the interaction. For 11 complexes, each dove gains $\mathcal{B}/2$. From Eq. \ref{sigma} and for $u < 1/2$, the $\sigma$-condition is written as

\be
\frac{\mathcal{B}}{\mathcal{C}} > \frac{1- u}{2u}
\ee

\nd Therefore, we have

\be
u_{1} = \frac{1}{ 1 + 2\mathcal{B}/\mathcal{C}}
\label{hawk-uc}
\ee

As discussed before we expect the evolutionarily stable strategy (ESS) to dominate for small values of the mutation rate $u < u_{1}$.
In the hawk-dove game, as a result of the multicellular population structure, the ESS is type 1 (dove). This can also be seen from numerical solutions of the model. The total type 0 and 1 abundances for various intensities of  selection are plotted in Fig. \ref{xtot_uvar}  for $\mathcal{B}=5$ and $\mathcal{C}=1$. For these benefit and cost values we observe $u_{1} = 0.09$ and $u_{2}=0.5$. We also plot the abundances of single cells and complexes  in Fig. \ref{Allx_uvar} for $w=0.1$. In Fig. \ref{gm-phase}  the phase diagram of the hawk-dove game is plotted in agreement with Eq. \ref{hawk-uc}. The regions between the curve $(u_{1},\mathcal{B}/\mathcal{C})$ and $(u_{2},\mathcal{B}/\mathcal{C})$ are where the hawk strategy is favored and vice versa. The topology of this phase diagram is different from the cooperation game as the two phase boundaries $(u_{1}, \mathcal{B}/\mathcal{C})$ and $(u_{2},\mathcal{B}/\mathcal{C})$ meet at $\mathcal{B}=2\mathcal{C}$.

\subsection{Average fitness}
We now study how the average fitness at equilibrium, $\phi^{\star}$, depends on the mutation rate, $u$. The average fitness at equilibrium is calculated in Appendix A, Eq.\ref{quad-eta-phi}
\be
\phi^{\star} = -\frac{1}{2} +\frac{1}{2}\sqrt{1 + 4\big(D_{0}(1-z)+D_{1}z\big)}
\label{phi-ave}
\ee

\nd Here $z\equiv \eta^{\star}/(1+\eta^{\star})=x^{\star}_{1}/(x^{\star}_{0}+x^{\star}_{1})$ is the fraction of type 1 singlets. Note that $z$ is a function of $u$ and of the payoff values. From Eq. \ref{ABC}, the coefficients $D_{0}$ and $D_{1}$ are

\bea
D_{0} &=& 2\big( 1 + wa\big) + w\big( b + c - 2a\big)u\nonumber\\ 
D_{1} &=& 2\big( 1 + wd\big) + w\big( b + c - 2d\big)u
\eea

If $a > d$ then type 0 is ESS and as $u \to 0$, we have $z \to 0$. In this case $z$ is an increasing function near $u=0$ and therefore $z^{'}(u=0) >0$. If $a < d$ then type 1 is ESS and as $u \to 0$, we have $z \to 1$. In this case $z^{'}(u=0) < 0$. Denoting average fitness in this limit with $\phi^{\star}_{0}$ we have

\be
\phi^{\star}_{0}=  -\frac{1}{2} +\frac{1}{2}\sqrt{9 + 8w\cdot{\rm max}(a, d)} \nonumber\\
\ee

\nd Similarly one can show that at $u = 1$ the average fitness, denoted by $\phi^{\star}_{1}$, is

\be
\phi^{\star}_{1}= -\frac{1}{2} +\frac{1}{2}\sqrt{9 + 4w(b+c)}
\ee

\begin{figure}[h!]
\begin{center}
\epsfig{figure=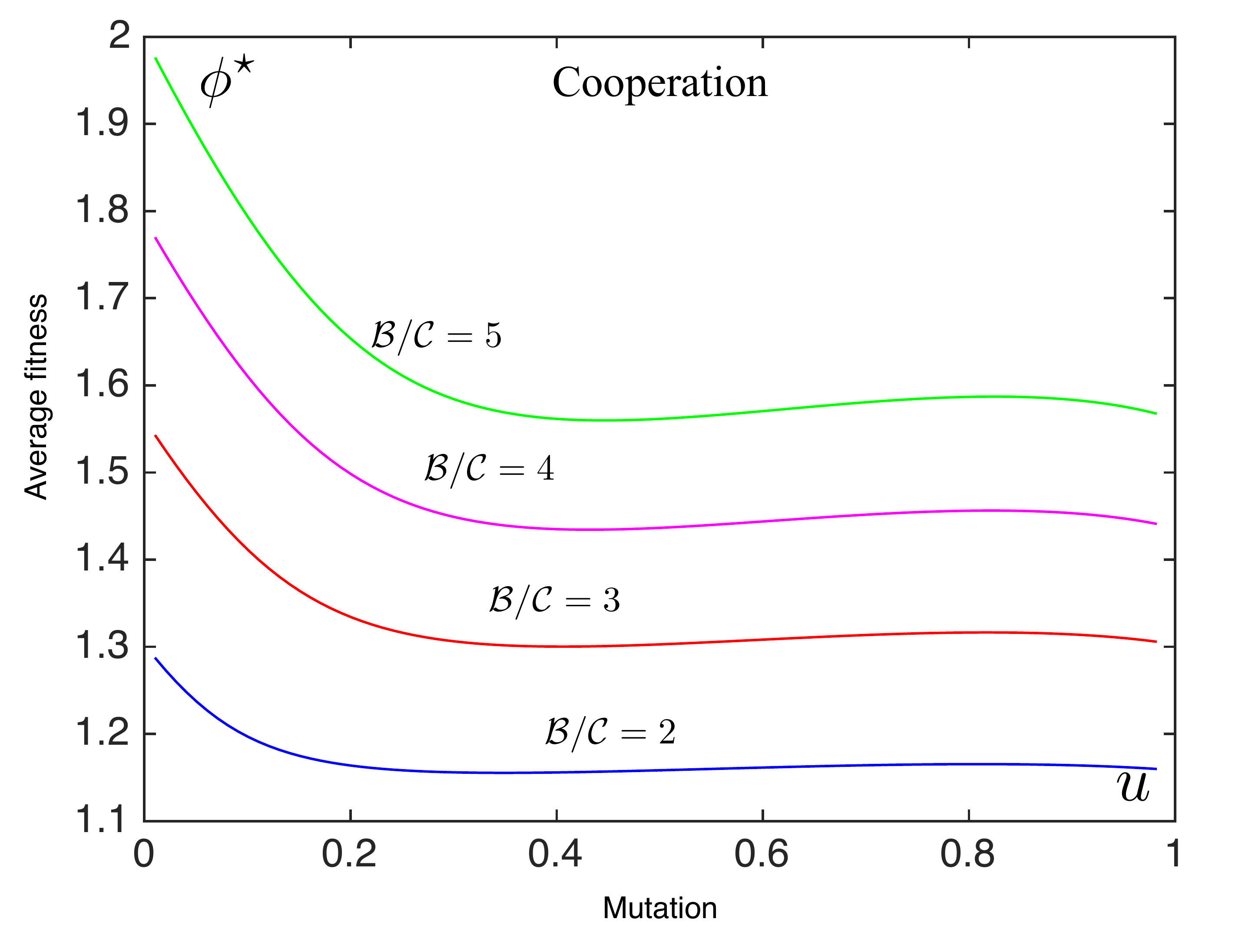, height=190pt,width=240pt,angle=0}
\epsfig{figure=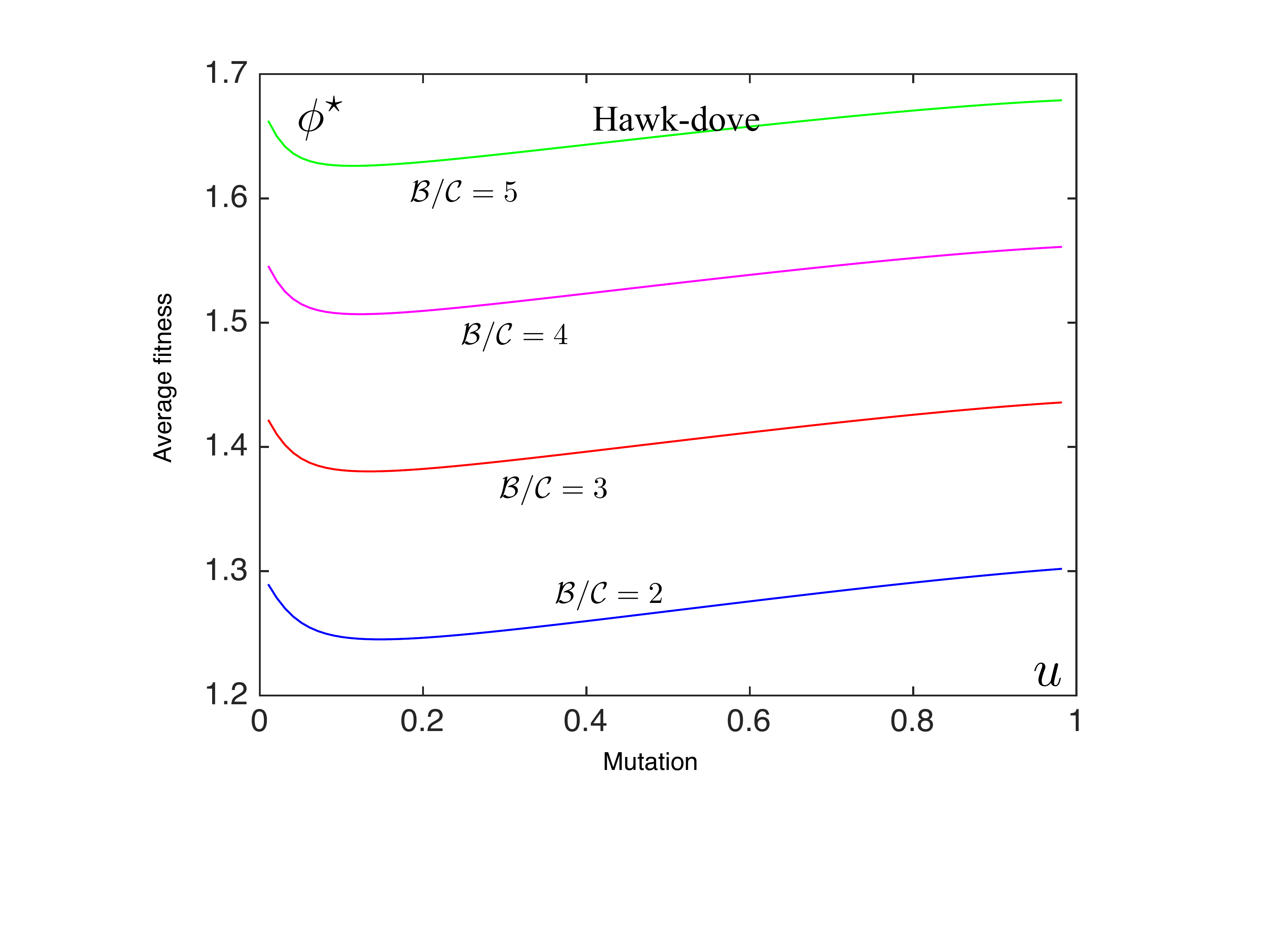, height=190pt,width=240pt,angle=0}
\end{center}
\caption{Numerical solutions for the equilibrium average fitness, $\phi^{\star}$, as a function of the mutation rate, $u$. For the game of cooperation $\phi^{\star}$ has a global maximum at $u=0$ and a local minimum at $u=1$ (top). For the hawk-dove game, $\phi^{\star}$ has a local maximum at $u=0$ and a global maximum at $u=1$ (bottom).
Parameter values: $w=0.5$ and $\mathcal{C}=1$.}
\label{phi-ave-plot}
\end{figure}

For the game of cooperation, if the benefit-to-cost ratio is greater than unity, the condition $(b + c)/2 < a = {\rm max}(a, d)$ is satisfied. Thus $\phi^{\star}_0$ is larger than  $\phi^{\star}_{1}$. For the hawk-dove game, on the other hand, we have $(b + c)/2 < d = {\rm max}(a,d)$ and thus $\phi^{\star}_{0} < \phi^{\star}_{1}$.

We now show that for the hawk-dove game, $\phi^{\star}_{0}$ and $\phi^{\star}_{1}$ are local and global maxima of $\phi^{\star}$, respectively.
Since $\phi^{\star}$ is an increasing function of $G \equiv D_{0}(z-1) + D_{1}z$, we look at derivative of this term as a function of $u$ 

\bea
\frac{{\rm d}G}{{\rm d}u} &=& D^{'}_{0} + \big(D^{'}_{1}-D^{'}_{0}\big)z + \big(D_{1}-D_{0}\big)z^{'}\nonumber\\
&=& w(b+c-2a) + 2w(a-d)z- 2w(a-d)(1-u)z^{'}
\eea

\nd The prime symbol represents the derivative with respect to $u$. At $u=0$ we have $z(u=0)=1$ and $z^{'}(u=0) < 0$. For $u=1$ we have $z(u=1) < 1$ and $z^{'}(u=1) > 0$. Using these results and inserting the payoff values of the hawk-dove game we obtain

\bea
\frac{{\rm d}G}{{\rm d}u}\at{u=0}&=& w\mathcal{C}\cdot z^{'}(u=0) < 0\nonumber\\
\frac{{\rm d}G}{{\rm d}u}\at{u=1} &=& w\mathcal{C} \big( 1- z(u=1)\big) > 0
\label{deriv-hd}
\eea

\nd The inequalities are true for $\mathcal{C} > 0$. Thus for the hawk-dove, $\phi^{\star}_{0}$ and $\phi^{\star}_{1}$ are local maxima. The average fitness at $u=1$ mutation is a global maximum. The signs of derivatives at
Eq. \ref{deriv-hd} imply that there is a global minimum for $u^{\star}$ at $0 < u^{\star} < 1/2$. These results can be compared with numerical results in Fig. \ref{phi-ave-plot}.

The same analysis can be done for the game of cooperation. Following similar steps we get

\bea
\frac{{\rm d}G}{{\rm d}u}\at{u=0}&=& -w\big(\mathcal{B}-\mathcal{C}\big)\big(1+2z^{'}(u=0)\big) < 0\nonumber\\
\frac{{\rm d}G}{{\rm d}u}\at{u=1} &=& -w\big(\mathcal{B}-\mathcal{C}\big)\big(1-2z(u=1)\big) < 0
\label{deriv-coop}
\eea

\nd The inequalities hold for $\mathcal{B} > \mathcal{C}$. We have also used $z(u=0)=0, z^{'}(u=0) > 0$ and $z(u=1) < 1/2, z^{'}(u=1) <0$.

\section{Model for larger complexes, $n \geq 3$}

\begin{figure}
\begin{center}
\epsfig{figure=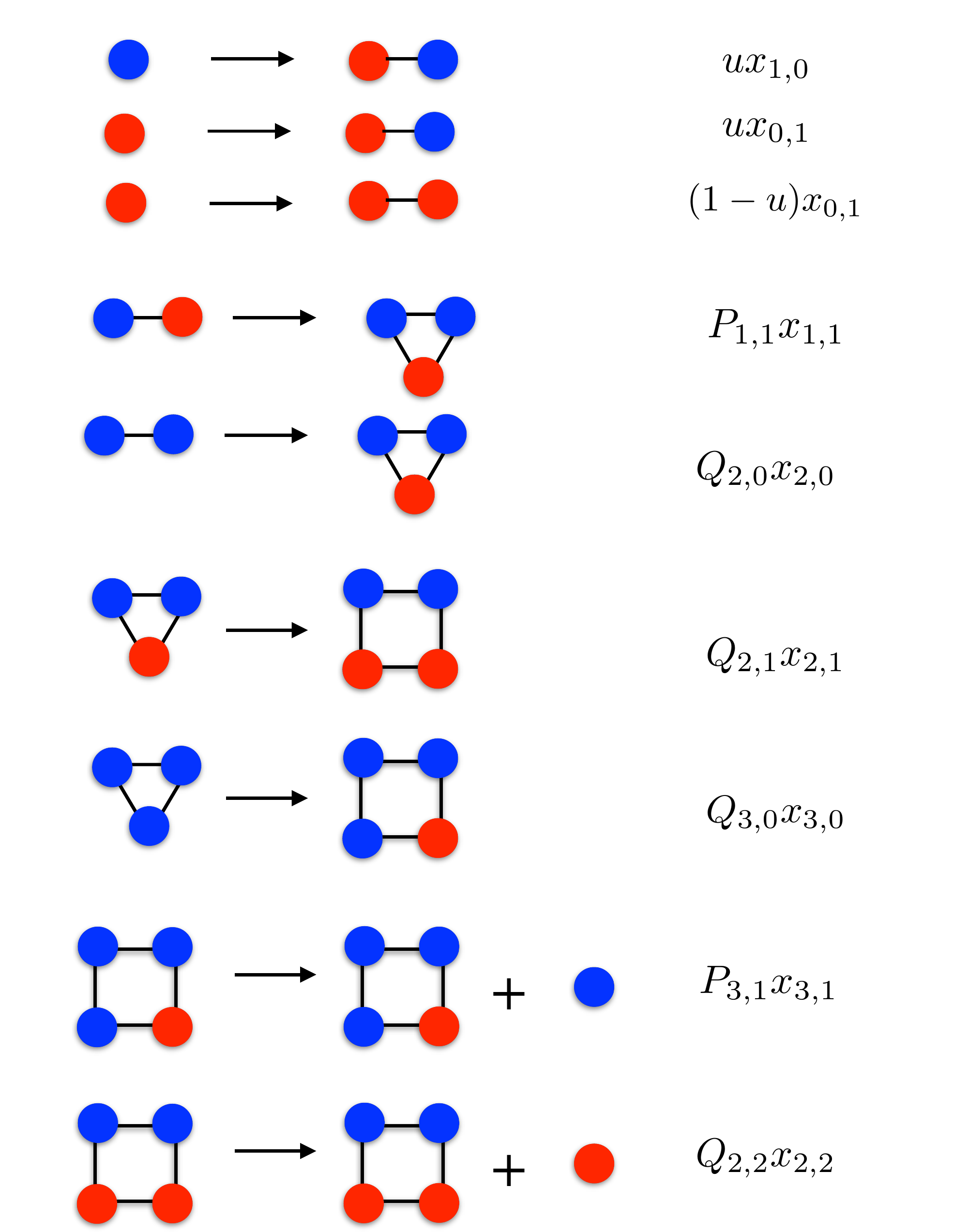, height=250pt,width=210pt,angle=0}
\end{center}
\caption{A general model for games of multicellularity. The figure depicts some of the possible reactions and the corresponding rates for $n=4$. The abundance of a complex of size $k$ with $i$ many type 0 cells is denoted by $x_{i,k-i}$. The coefficients $P_{i,j}$ denote the rates at which complexes with $i$ type 0 cells and $j$ type 1 cells generate type 0 offsprings. Similarly, $Q_{i,j}$ denote the rate at which the same complex generates type 1 cells. (See Eq. \ref{P-gen}.)}
\label{reaction2}
\end{figure}

We now generalize the model to arbitrary maximum complex size, $n$.  Complexes of sizes 2 to  $n$ can now coexist. Each complex can be a mixture of type 0 and type 1 cells. Cells divide and offspring stay together with their parents. Thus, a complex of size $k$ produces a complex of size $k+1$. Each offspring adopts its parent's type with probability $1-u$ and changes to the other type with probability $u$. The offspring of cells inside a complex of maximum size leave the complex and the pool of single cells. Cells inside a complex gain payoff through a biological game with pairwise interactions. 

The abundance of a complex of size $k$ with $i$ many cells of type 0 cells and $k-i$ many cells of type 1 is denoted by $x_{i,k-i}$. For example, $x_{00}$ in previous notation is now $x_{2,0}$. Similarly, $x_{01}$ becomes $x_{1,1}$, while $x_{11}$ becomes $x_{0,2}$ and so on. 
 
The evolutionary dynamics can be written as a system of differential equations

\bea
\dot{x}_{1,0} &=& \sum^{n}_{l=0}P_{l,n-l}x_{l,n-l} -x_{1,0} -x_{1,0}\phi\nonumber\\
\dot{x}_{0,1}&=& \sum^{n}_{l=0}Q_{l,n-l}x_{l,n-l} -x_{0,1} -x_{0,1}\phi\nonumber\\
&\vdots&\nonumber\\
\dot{x}_{i,k-i}&=& P_{i-1,k-i}x_{i-1,k-i}+Q_{i,k-i-1}x_{i,k-i-1}\nonumber\\
&-&\Big(P_{i,k-i}+Q_{i,k-i}\Big)x_{i,k-i}-x_{i,k-i}\phi\nonumber\\
&&(1\leq k < n, 0 \leq i \leq k)\nonumber\\
&\vdots&\nonumber\\
\dot{x}_{i,n-i}&=& P_{i-1,n-i}x_{i-1,n-i}+Q_{i,n-i-1}x_{i,n-i-1}\nonumber\\
&-&x_{i,n-i}\phi\nonumber\\
&&(0 \leq i \leq n)
\label{ode-gen}
\eea

\nd Again the average fitness $\phi$ is given by the condition that total relative abundances of type 0 and type 1 cells add up to one, that is

\be
\sum^{n}_{k=1}k\cdot \Big(\sum^{k}_{i=0}x_{i,k-i} \Big)= 1
\label{phi-cons}
\ee

\nd Thus,

\be
\phi= \sum^{n}_{k=1}\sum^{k}_{i=0} \Big( P_{i,k-i}+Q_{i,k-i}\Big)x_{i,k-i}
\label{phi-gen}
\ee

\nd The coefficients $P_{i,j}$ and $Q_{i,j}$ are the production rates for creating a type 0 or type 1 cell inside a complex of size $k=i+j$. They are expressed in terms of the game payoffs, the mutation rate, $u$, and the intensity of selection, $w$

\bea
P_{i,j} &=& i\cdot\big(1+w(a(i-1)+bj)\big)(1-u) + j\cdot\big(1+w(ic+(j-1)d)\big)u\nonumber\\
Q_{i,j}&=& i\cdot\big(1+w(a(i-1)+bj)\big)u + j\cdot\big(1+w(ic+(j-1)d)\big)(1-u)\nonumber\\
\label{P-gen}
\eea

\nd Here $i$ is the number of type 0 cells and $j=k-i$ is the number of type 1 cells inside the complex. 

The fixed points of the above system of equations are obtained by putting the right-hand side of Eq. \ref{ode-gen} equal to zero. The solutions, $x^{\star}_{i,k-i}$ are given by

\bea
x^{\star}_{1,0}&=&\frac{1}{1+\phi^{\star}}\sum^{n}_{l=0}P_{l,n-l}x^{\star}_{l,n-l}\nonumber\\
x^{\star}_{0,1}&=&\frac{1}{1+\phi^{\star}}\sum^{n}_{l=0}Q_{l,n-l}x^{\star}_{l,n-l}\nonumber\\
&\vdots&\nonumber\\
x^{\star}_{i,k-i} &=& \Big( \phi^{\star} + P_{i,k-i}+Q_{i,k-i}\Big)^{-1}\nonumber\\
&\times& \Big( P_{i-1,k-i}x^{\star}_{i-1,k-i}+Q_{i,k-i-1}x^{\star}_{i,k-i-1}\Big)\nonumber\\
&&(0<k<n , 0 \leq i \leq k)\nonumber\\
&\vdots&\nonumber\\
x^{\star}_{i,n-i}&=&\frac{1}{\phi^{\star}}\Big( P_{i-1,n-i}x^{\star}_{i-1,n-i}+Q_{i,n-i-1}x^{\star}_{i,n-i-1}\Big)\nonumber\\
&&(0 \leq i \leq n)
\label{recurr1}
\eea

To solve Eq. \ref{recurr1} analytically, we can, in principle, follow the same method as used in Section 2 (and in Appendix A). The abundances  $x^{\star}_{i,k-i}$ can be expressed in terms of $\eta^{\star}\equiv x^{\star}_{0,1}/x^{\star}_{1,0}$ and the average fitness $\phi^{\star}$. The system of equations \ref{recurr1} can be reduced to two equations for
$\eta^{\star}$ and $\phi^{\star}$. We have presented a sketch of this method in Appendix C for $n=3$. A general approach for arbitrary $n$ is similar but the solutions become cumbersome as $n$ increases. 

At $w=0$ the dynamics is neutral. In this limit we have $\phi^{\star}=1$. Eq. \ref{recurr1} becomes a system of linear recurrence equations

\bea
\hat{x}^{\star}_{1,0}&=&\frac{1}{2}\sum^{n}_{l=0}\hat{P}_{l,n-l}\hat{x}^{\star}_{l,n-l}\nonumber\\
\hat{x}^{\star}_{0,1}&=&\frac{1}{2}\sum^{n}_{l=0}\hat{Q}_{l,n-l}\hat{x}^{\star}_{l,n-l}\nonumber\\
&\vdots&\nonumber\\
\hat{x}^{\star}_{i,k-i} &=& \frac{1}{k+1}\Big( \hat{P}_{i-1,k-i}\hat{x}^{\star}_{i-1,k-i}+\hat{Q}_{i,k-i-1}\hat{x}^{\star}_{i,k-i-1}\Big)\nonumber\\
&&(1\leq k < n, 0 \leq i \leq k)\nonumber\\
&\vdots&\nonumber\\
\hat{x}^{\star}_{i,n-i}&=&\Big( \hat{P}_{i-1,n-i}\hat{x}^{\star}_{i-1,n-i}+\hat{Q}_{i,n-i-1}\hat{x}^{\star}_{i,n-i-1}\Big)\nonumber\\
&&(0 \leq i \leq n)
\label{recurr2}
\eea

\nd Here $\hat{x}$ denotes the $w=0$ limit. Similarly, $\hat{P}_{i,j}=(1-u)i+uj$ and $\hat{Q}_{i,j}=ui+(1-u)j$ are zero selection limits of $P_{i,j}$ and $Q_{i,j}$. The solutions $\hat{x}_{i,j}$ have the following properties: (i) $x^{\star}_{i,j} = x^{\star}_{j,i}$. 
(ii) The frequencies of type 0 and type 1 cells are the same: $\sum_{i,k} i x^{\star}_{i,k-i} = \sum_{i,k} i x^{\star}_{k-i,i}=1/2$.
(iii) The total abundance of cells in complexes of the same size, $k$, is independent of the mutation rate.

\section{The $\sigma$-condition for $n\geq 2$}
In this section we derive the $\sigma$-condition for the generalized model introduced in previous section at the weak selection limit. 
For $n=2$, we showed that  $\sigma_{2}(u)=(1-u)/u$. It turns out that for $n > 2$ the parameter $\sigma_{n}$ in Eq. \ref{sigma0}, is now given by a ratio of two polynomials of degree $n-1$ in $u$. In fact the general form for $\sigma_{n}(u)$ is $(c_{n}-h_{n}(u))/h_{n}(u)$. Here $h_{n}(u)$ is a polynomial in $u$, and $c_{n}$ is a numerical constant. The symmetry of the $\sigma$-condition is preserved as we increase the maximum complex size. We will later compare numerical solutions for strategy selection in the general model and analytical prediction of $\sigma$-condition. We observe that our predicted $\sigma$-condition holds well above the weak selection limit as well.

We rewrite Eq. \ref{recurr1} in the following form

\bea
x^{\star}_{1,0}&=&\frac{1}{\phi^{\star}}\Big(\sum^{n}_{l=0}P_{l,n-l}x^{\star}_{l,n-l} - x^{\star}_{1,0}\Big)\nonumber\\
x^{\star}_{0,1}&=&\frac{1}{\phi^{\star}}\Big(\sum^{n}_{l=0}Q_{l,n-l}x^{\star}_{l,n-l} - x^{\star}_{0,1}\Big)\nonumber\\
&\vdots&\nonumber\\
x^{\star}_{i,k-i} &=& \frac{1}{\phi^{\star}}\Big\{\Big( P_{i-1,k-1}x^{\star}_{i-1,k-i}+Q_{i,k-i-1}x^{\star}_{i,k-i-1}\Big)\nonumber\\
&-& \Big(P_{i,k-i}+Q_{i,k-i}\Big)x^{\star}_{i,k-i}\Big\}\nonumber\\
&&(1\leq k < n, 0 \leq i \leq k)\nonumber\\
&\vdots&\nonumber\\
x^{\star}_{i,n-i}&=&\frac{1}{\phi^{\star}}\Big( P_{i-1,n-i}x^{\star}_{i-1,n-i}+Q_{i,n-i-1}x^{\star}_{i,n-i-1}\Big)\nonumber\\
&&(0 \leq i \leq n)
\label{recurr3}
\eea

Total number of type 0 and type 1 cells are written in terms of solutions of Eq. \ref{recurr3} as

\bea
x^{\star}_{\rm tot,0} &=& \sum^{n}_{k=1}\sum^{k}_{i=0} i \cdot x^{\star}_{i,k-i}\nonumber\\
x^{\star}_{\rm tot,1}&=& \sum^{n}_{k=1}\sum^{k}_{i=0} (k-i) \cdot x^{\star}_{i,k-i}
\label{xtot-gen}
\eea

The condition for type 0 selection, Eq. \ref{xtot-01}, is $x^{\star}_{\rm tot,0} > x^{\star}_{\rm tot,1}$. To have a first order estimate of abundances 
in powers of selection intensity we substitute $x_{i,j}$ in left-hand side of Eq. \ref{recurr3} with $w=0$ limit solutions, $\hat{x}_{i,j}$. 
This way $x_{i,k-i}$ from left-hand side of Eq. \ref{recurr3} is expressed to zeroth and first order in $w$. Substituting  solutions into Eq. \ref{xtot-gen} and Eq. \ref{xtot-01} gives rise to a closed form generalized $\sigma$-condition.  After some straightforward algebra we get

\be
\sum^{n}_{k=1}\sum^{k}_{i=0} P_{i,k-i}\hat{x}^{\star}_{i,k-i} > \sum^{n}_{k=1}\sum^{k}_{i=0} Q_{i,k-i}\hat{x}^{\star}_{i,k-i}
\label{gen-sigma0}
\ee

\nd This is a generalization of Eq. \ref{sigma-PQ-n2}. We can write this in terms of fitness gains of either of type 0 and 1 strategies as we did for $n=2$ case. Denoting fitness gains for type 0 and type 1 in weak selection by $\delta f_{0,n}$ and $\delta f_{1,n}$, respectively, we can write:

\bea
\delta f_{0,n} &=&\sum^{n}_{k=2}\sum^{k}_{i=1}i\big((i-1)a+(k-i)b\big)\cdot \hat{x}^{\star}_{i,k-i}\nonumber\\
\delta f_{1,n} &=&\sum^{n}_{k=2}\sum^{k}_{i=1}i\big((i-1)d+(k-i)c\big)\cdot \hat{x}^{\star}_{k-i,i}
\label{df-n}
\eea

\nd Then Eq. \ref{gen-sigma0} can be rewritten as

\be
\Big(u - \frac{1}{2}\Big)\big( \delta f_{0,n}  - \delta f_{1,n}\big) < 0
\label{sigma-df-n}
\ee

\nd The zeros of the term $\delta f_{0,n}-\delta f_{1,n}$ determine $u_{1}$ whereas $u_{2}$ is zero of $u - 1/2$. 

Let us consider $n=3$ as an example. In this case Eq. \ref{gen-sigma0} is written as 
\bea
P_{3,0}\hat{x}^{\star}_{3,0}+P_{2,1}\hat{x}^{\star}_{2,1}+P_{1,2}\hat{x}^{\star}_{1,2}+P_{0,3}\hat{x}^{\star}_{0,3}\nonumber\\
+P_{2,0}\hat{x}^{\star}_{2,0} + P_{1,1}\hat{x}^{\star}_{1,1}+P_{0,2}\hat{x}^{\star}_{0,2} + P_{1,0}\hat{x}^{\star}_{1,0}+P_{0,1}\hat{x}^{\star}_{0,1}\nonumber\\
> \nonumber\\
Q_{3,0}\hat{x}^{\star}_{3,0}+Q_{2,1}\hat{x}^{\star}_{2,1}+Q_{1,2}\hat{x}^{\star}_{1,2}+Q_{0,3}\hat{x}^{\star}_{0,3}\nonumber\\
+Q_{0,2}\hat{x}^{\star}_{0,2} + Q_{1,1}\hat{x}^{\star}_{1,1}+Q_{2,0}\hat{x}^{\star}_{2,0} + Q_{1,0}\hat{x}^{\star}_{1,0}+Q_{0,1}\hat{x}^{\star}_{0,1}
\label{sigma2}
\eea

\nd The terms corresponding to singlets $\hat{x}_{1,0}, \hat{x}_{0,1}$ can be dropped from both sides of the inequality since they do not confer any fitness gain or loss. Similarly Eq. \ref{sigma-df-n} is

\be
\Big(u-\frac{1}{2}\Big) \big( \delta f_{0,3}-\delta f_{1,3}\big) < 0
\label{sigma3}
\ee

\nd The fitness gains $\delta f_{0,3}$ and $\delta f_{1,3}$ are

\bea
\delta f_{0,3} &=& 2a\cdot \hat{x}^{\star}_{2,0} + b\cdot \hat{x}^{\star}_{1,1}\nonumber\\ 
&+& 6a \cdot \hat{x}^{\star}_{3,0} + 2\big(a+b\big)\cdot \hat{x}^{\star}_{2,1} + 2b \cdot \hat{x}^{\star}_{1,2}\nonumber\\
\delta f_{1,3}&=& 2d \cdot \hat{x}^{\star}_{0,2} + c \cdot \hat{x}^{\star}_{1,1}\nonumber\\ 
&+& 6a \cdot \hat{x}^{\star}_{0,3} + 2\big(d+c\big)\cdot \hat{x}^{\star}_{1,2} + 2c \cdot \hat{x}^{\star}_{2,1}\nonumber\\
\label{df3}
\eea

The $n=3$ solutions for Eq. \ref{recurr2} are

\bea
\hat{x}_{1,0}&=&\hat{x}_{0,1}=\frac{3}{22}\nonumber\\
\hat{x}^{\star}_{2,0,}&=& \hat{x}^{\star}_{3,0} = \frac{1-u}{22}\nonumber\\
\hat{x}^{\star}_{1,1}&=&  \frac{u}{11}\nonumber\\
\hat{x}^{\star}_{3,0}&=& \hat{x}^{\star}_{0,3} = \frac{(1-u)^{2}}{11}\nonumber\\
\hat{x}^{\star}_{2,1}&=& \hat{x}^{\star}_{1,2} = \frac{u(2-u)}{11}\nonumber\\
\label{recurr-n3}
\eea

Substituting the above results back into Eq. \ref{df3} we can write the $\sigma$-condition, Eq. \ref{sigma3}, as

\be
\Big( u - \frac{1}{2}\Big)\Big( (a-d) \frac{4u^{2}-9u+7}{-4u^{2}+9u} + (b-c)\Big) <  0
\label{sigma-n3}
\ee

\nd Now $u_{1}$ is the solution of $(a-d)(4u^{2}-9u+7)/(9u-4u^{2}) + (b-c)=0$. 

\begin{figure}
\begin{center}
\epsfig{figure=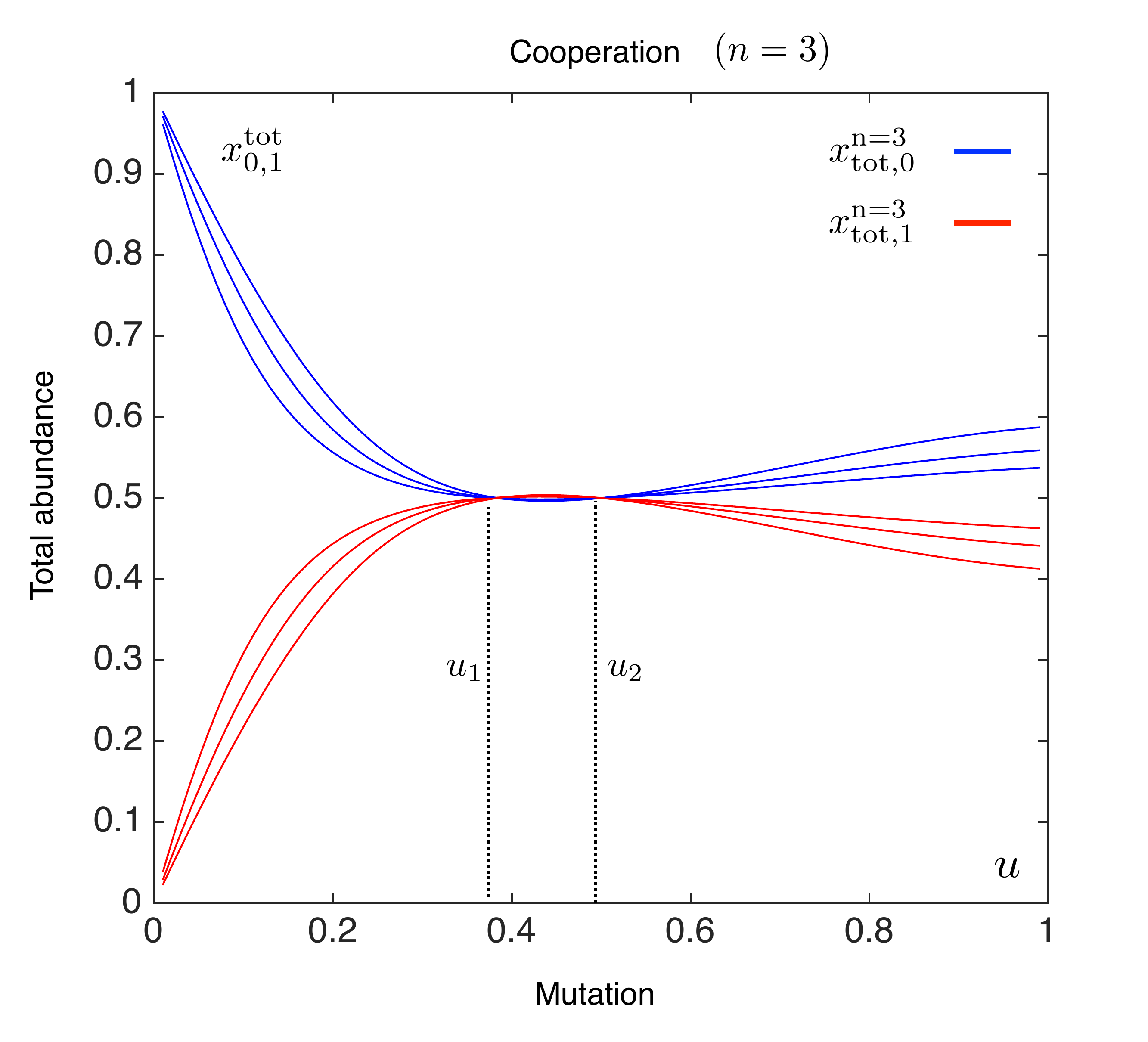, height=220pt,width=240pt,angle=0}
\epsfig{figure=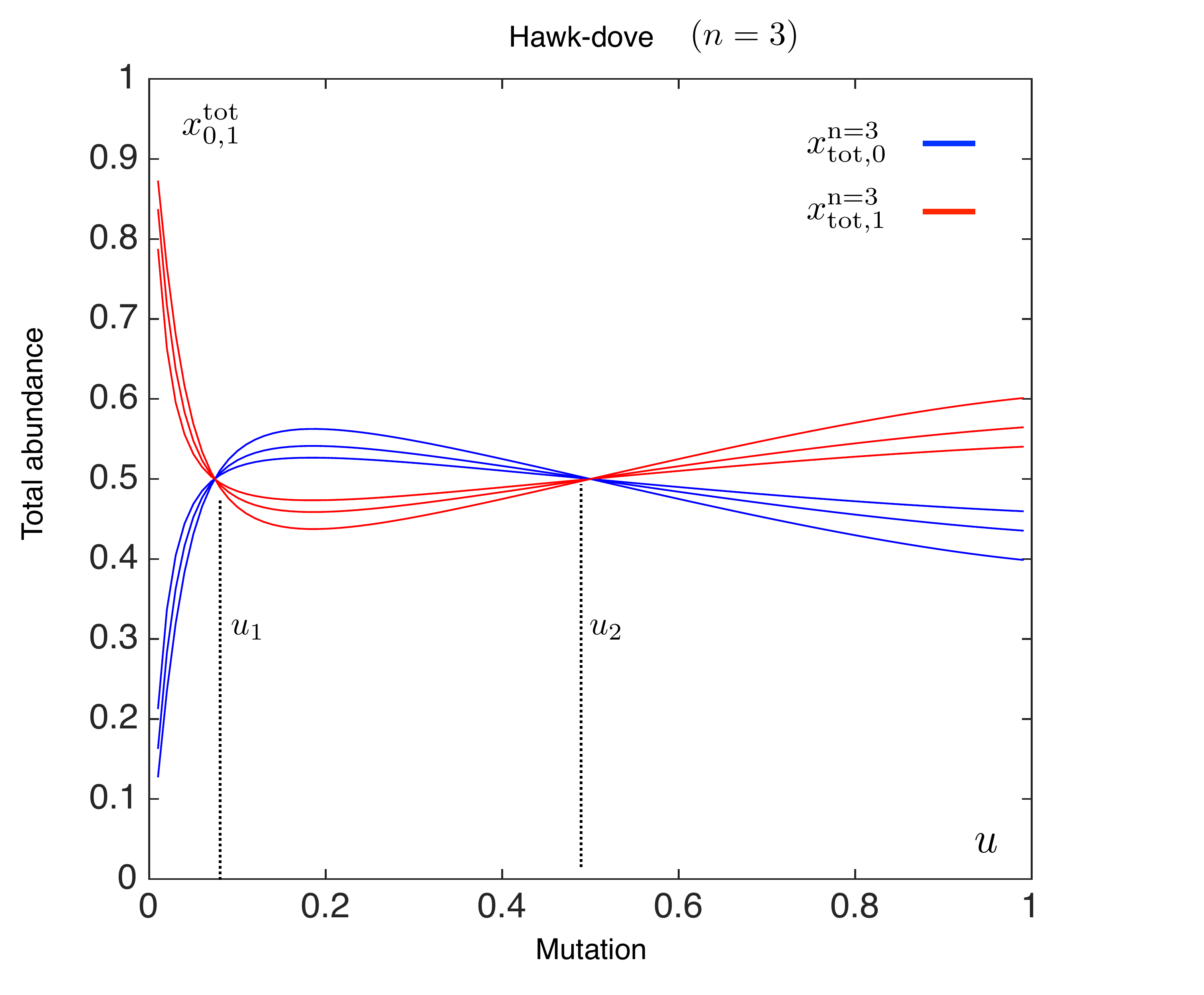, height=220pt,width=240pt,angle=0}
\end{center}
\caption{Numerical solutions of $x^{\star}_{\rm tot,0}$ and $x^{\star}_{\rm tot,1}$ for $n=3$ for the cooperation game (top) and the hawk-dove game (bottom). Total abundances of each strategy are plotted as a function of $u$. Parameters are $\mathcal{B}=5,\mathcal{C}=1$ and $w=0.1,0.2,0.5$ (the curvature increases with $w$.) Notice that the main difference between $n=2$ and $n=3$ is the slight change in the value of $u_{1}$. While for $n=2$ the value of $u_{1}$ was independent of the selection intensity, for $n=3$ it is weakly dependent on $w$, but this is almost impossible to see from these plots.}
\label{plot-3complex-2complex-coop-uvar}
\end{figure}

\begin{figure}
\begin{center}
\epsfig{figure=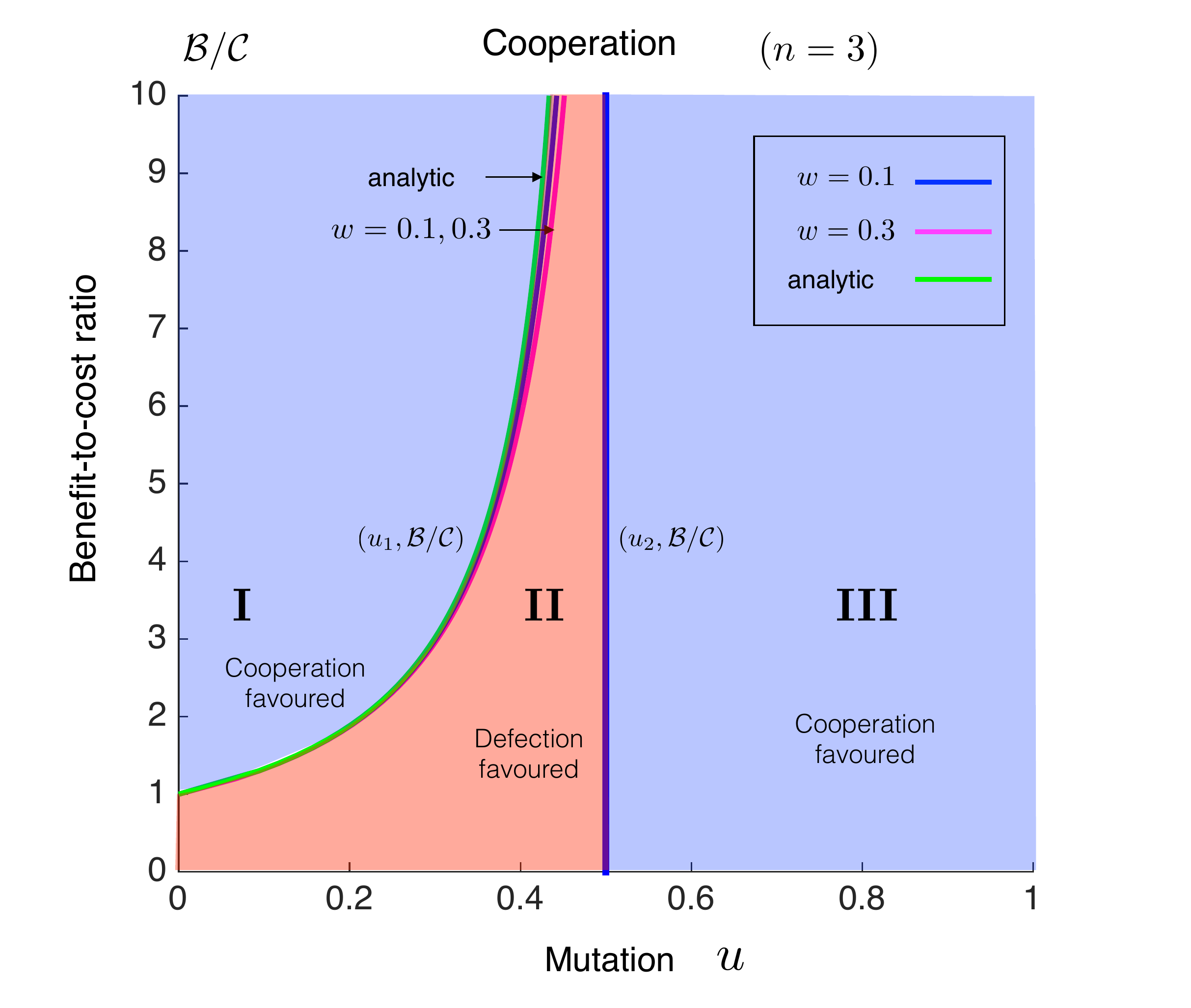, height=200pt,width=240pt,angle=0}
\epsfig{figure=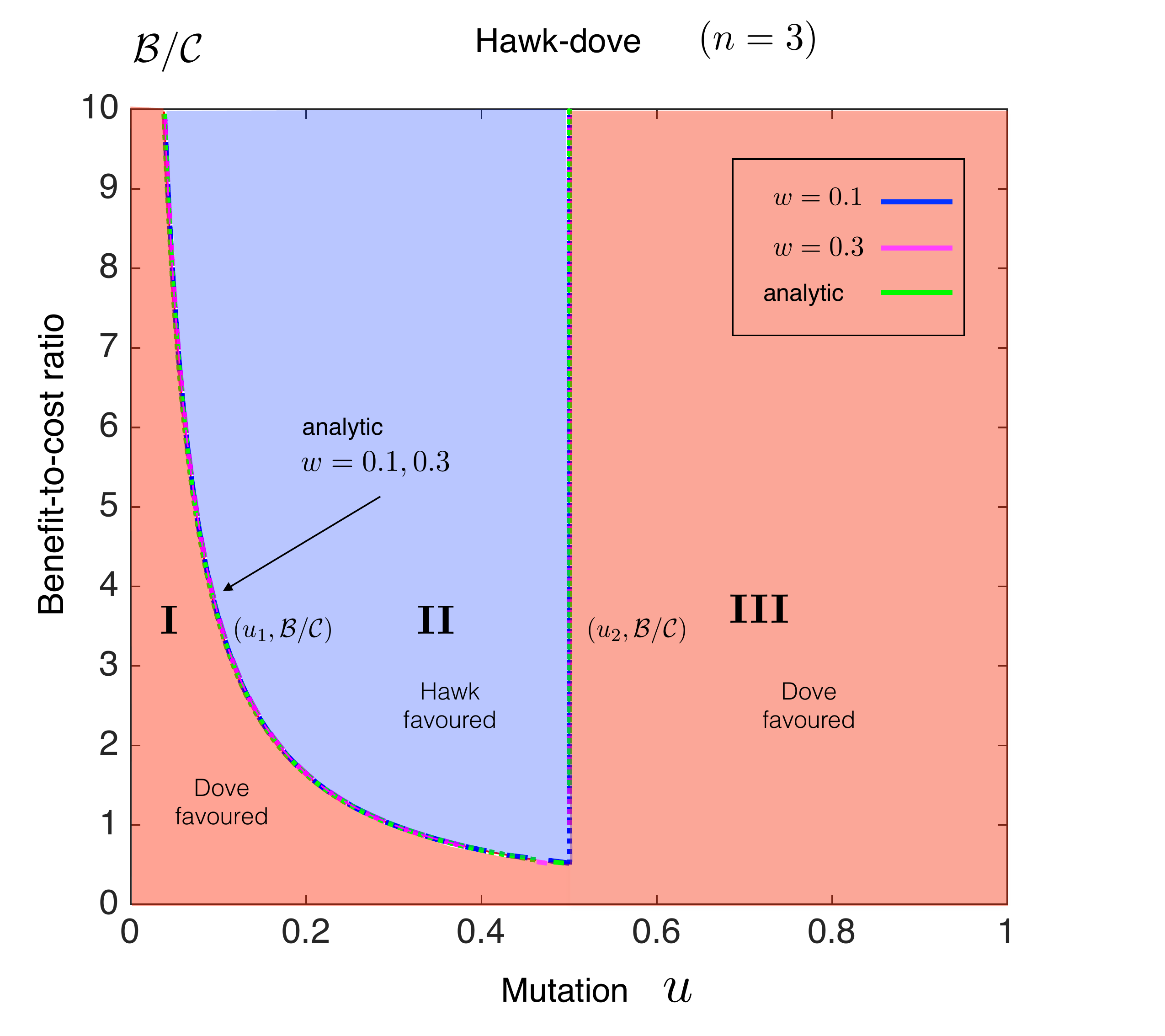, height=200pt,width=240pt,angle=0}
\end{center}
\caption{Phase diagram for the cooperation game (top) and the hawk-dove game (bottom) for maximum complex size, $n=3$. a) Cooperators are favored for $u < u_{1}$ and $u_{2} < u$ and $\mathcal{B} > \mathcal{C}$ (regions I and III). b) Hawks are favored for $u_{1} < u < u_{2}$ and $\mathcal{B} > 2\mathcal{C}$  (region II). As can be seen, the numerical and analytical results for phase boundaries are in excellent agreement.}
\label{gm-phase-n3}
\end{figure}

We have numerically solved the model for $n=3$ and
$\mathcal{B}=5$ and $\mathcal{C}=1$ for the cooperation game and the hawk-dove game. Predicted values for $u_{1}$ and $u_{2}$ are in excellent agreement for various selection intensities (Fig. \ref{plot-3complex-2complex-coop-uvar}). The phase diagrams for these games are depicted in 
Fig. \ref{gm-phase-n3} for $w=0.1$ and $0.3$. The phase boundaries match very well with the results from Eq. \ref{sigma-n3}. 

The same method can be used for larger maximum complex sizes at weak selection. Substituting for $\hat{x}_{i,k-i}$ from solutions of Eq. \ref{recurr2} into Eq. \ref{df-n} we can write

\bea
\delta f_{0,n} &=& \big( c_{n} - h_{n}(u) \big) a + h_{n}(u) b\nonumber\\
\delta f_{1,n} &=&  h_{n}(u) c + ( c_{n} - h_{n}(u) \big)d 
\eea

\nd From Eq. \ref{df-n}, the polynomial $h_{n}(u)$ and the constant $c_{n}$ are expressed in terms of $\hat{x}_{i,k-i}$ solutions

\bea
h_{n}(u) &=& \sum^{n}_{k=1}\sum^{k}_{i=0} i(k-i) \hat{x}^{\star}_{i,k-i}\nonumber\\
c_{n} &=& \sum^{n}_{k=1}\sum^{k}_{i=0}i(k-1)\hat{x}^{\star}_{i,k-i}
\eea

Using properties of Eq. \ref{recurr2} solutions, we can show that $c_{n}$ is in fact a constant. For $n=2$, we have $c_{2}(u)=1$ and $h_{2}(u)=u$. For $n=3$, from Eq. \ref{sigma-n3} we have $c_{3} = 7$ and $h_{3}(u)=-4u^{2}+9u$ (up to a constant common factor). Repeating the same steps for higher complex sizes we have computed $\sigma_{n}$ for $n=2,\ldots,20$. To avoid long formulas, below we write  $\sigma_{n}$ for only $n=2,\ldots,7$

\bea
\sigma_{2} &=&\frac{1-u}{u}\nonumber\\
\sigma_{3} &=&\frac{4u^{2}-9u+7}{-4u^{2}+9u}\nonumber\\ 
\sigma_{4} &=&\frac{-8u^{3}+28u^{2}-35u+23}{8u^{3}-28u^{2}+35u}\nonumber\\
\sigma_{5} &=&\frac{32u^{4}-160u^{3}+308u^{2}-281u+163}{-32u^{4}+160u^{3}-308u^{2}+281u}\nonumber\\
\sigma_{6} &=&\frac{-64u^{5}+432u^{4}-1176u^{3}+1640u^{2}-1215u+639}{64u^{5}-432u^{4}+1176u^{3}-1640u^{2}+1215u}\nonumber\\
\sigma_{7} &=&\frac{256u^{6}-2240u^{5}+8160u^{4}+-5968u^{3}+17980u^{2}-11439u+5553}{-256u^{6}+2240u^{5}-8160u^{4}+15968u^{3}-17980u^{2}+11439u}
\nonumber\\
\label{sigma-n}
\eea

\begin{figure}
\begin{center}
\epsfig{figure=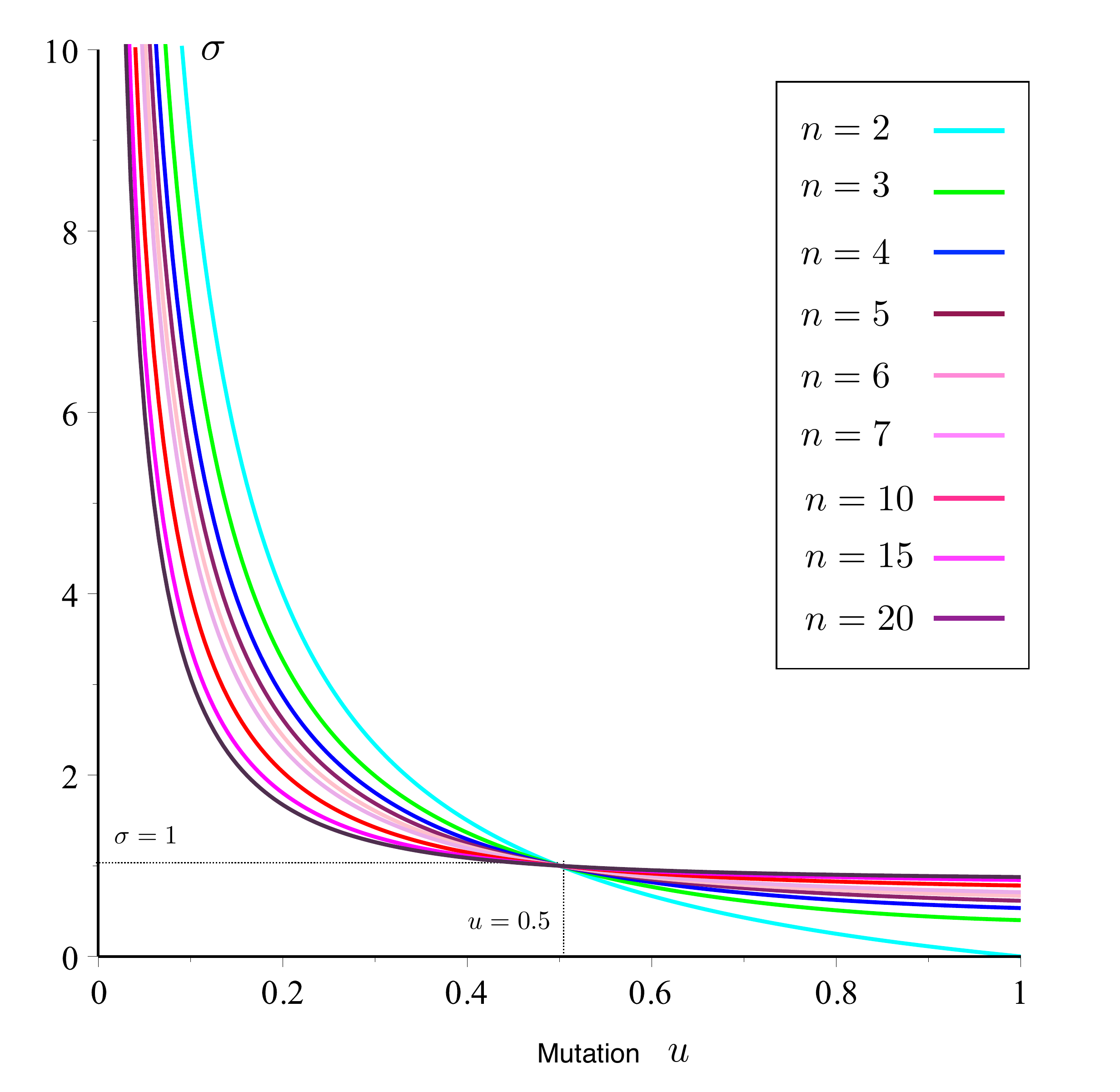, height=210pt,width=250pt,angle=0}
\end{center}
\caption{The value of $\sigma$ versus the mutation rate, $u$, is shown for various maximum complex sizes $n=2,3,4,5,6,7,10,15,20$.}
\label{plot-sigma}
\end{figure}

The results are plotted for $n=2,\ldots,20$ in Fig. \ref{plot-sigma}. As can be seen from above results, $h_{n}(u)$ always has a zero at $u=0$. Also $c_{n}$ is always a positive constant. The value of the critical mutation rate, $u_{1}$, is obtained from the equality

\be
\sigma_{n}(u_{1}) = \frac{c-b}{a-d}
\label{uc1-n}
\ee
 
\nd which can be solved for $u_{1}$ using Eq. \ref{sigma-n}. These estimated values of $u_{1}$ are compared with numerical solutions of the model
for various $n$, for the game of cooperation (Tabel \ref{table1}) and the hawk-dove game (Table \ref{table2}). Numerical values of $u_{1}$ are calculated for two selection intensities, $w=0.01$ and $w=0.1$ up to $n=7$ while theoretical estimate of $u_{1}$, Eq. \ref{uc1-n}, is presented up to $n=20$. 

We can now ask, for the general model, if the condition for evolutionary stability of a strategy is the same as in the case $n=2$, which is $a > d$. Proving a strategy is ESS by calculating the eigenvalues of the Jacobian is a cumbersome procedure for arbitrary $n$. Instead we assume that limit of $u \to 0$ of $\sigma$-condition lead to type 0 being ESS. We proved this in Appendix B for $n=2$ in weak selection. For $u \to 0$ and $2 \leq n \leq 7$, $\sigma_{n}$'s from Eq. \ref{sigma-n} have the limiting forms

\begin{table}[h!]
    \centering
    \begin{tabular}{cccccccccc}
        \hline
        \rowcolor[gray]{0.8}max. size, $n$&&&$u_{c,\rm analytic}$&&&$u_{c,\rm num}(w=0.01)$&&$u_{c,\rm num}(w=0.1)$\\\hline
        2&&&0.400&&&0.400&&0.400\\\hline
         3&&& 0.373&&&0.373&&0.377\\\hline
          4&&&0.352&&& 0.353&&0.359\\\hline
           5&&&0.335&&&0.336&&0.346\\\hline
            6&&& 0.321&&&0.325&&0.322\\\hline
             7&&& 0.309&&&0.306&&0.312\\\hline
10&&& 0.281&&&-&&-\\\hline
    15&&& 0.251&&&-&&-\\\hline
     20&&& 0.231&&&-&&-\\\hline
    \end{tabular}
\caption{The critical mutation rate, $u_{1}$, for the game of cooperation. Cooperators are favored if $u<u_{1}$.  Parameter values are $\mathcal{B}=5$ and $\mathcal{C}=1$. Numerical solutions for $w=0.01$ and $w=0.1$ are shown up to $n=7$.}
\label{table1}
\end{table}

\begin{table}[h!]
    \centering
    \begin{tabular}{cccccccccc}
        \hline
        \rowcolor[gray]{0.8}max. size, $n$&&&$u_{c,\rm analytic}$&&&$u_{c,\rm num}(w=0.01)$&&$u_{c,\rm num}(w=0.1)$\\\hline
        2&&&0.090&&&0.090&&0.090\\\hline
         3&&& 0.073&&&0.073&&0.073\\\hline
          4&&&0.063&&&0.063&&0.063\\\hline
           5&&&0.056&&&0.056&&0.057\\\hline
            6&&& 0.051&&&0.053&&0.053\\\hline
             7&&& 0.047&&&0.042&&0.049\\\hline
   10&&& 0.040&&&-&&-\\\hline
   15&&&0.034&&&-&&-\\\hline
    20&&& 0.029&&&-&&-\\\hline
    \end{tabular}
\caption{The critical mutation rate, $u_{1}$, for the hawk-dove game. Doves are favored for $u<u_{1}$. Parameter values are $\mathcal{B}=5$ and $\mathcal{C}=1$. Numerical solutions for $w=0.01$ and $w=0.1$ are shown up to $n=7$.}
\label{table2}
\end{table}

\bea
\sigma_{2}(u \to 0)&=& \frac{1-u}{u}\nonumber\\
\sigma_{3}(u \to 0)&\sim& \frac{(7/9)-u}{u}\nonumber\\
\sigma_{4}(u \to 0)&\sim& \frac{(23/35)-u}{u}\nonumber\\
\sigma_{5}(u \to 0)&\sim& \frac{(163/281)-u}{u}\nonumber\\
\sigma_{6}(u \to 0)&\sim& \frac{(639/1215)-u}{u}\nonumber\\
\sigma_{7}(u \to 0)&\sim& \frac{(5553/11439)-u}{u}\nonumber\\
\eea

\nd The leading term coefficients $1, 7/9, 23/35, \ldots$ are all positive and thus in the limit $u \to 0$ lead to $a > d$. 

Substituting the above results into Eq. \ref{uc1-n}, one can see that no matter how large $n$ is, the value of $u_{1}$ always remains positive. In other words, the phase boundary $(u_{1},\mathcal{B}/\mathcal{C})$ does not hit the $y$-axis for some large-$n$. If type 0 is ESS for a given game at a small value of $n$, it will remain ESS also for large $n$. Interestingly the plot of sigma versus the mutation rate also represents part of the phase boundary in the hawk- dove game. From Eq. \ref{uc1-n} and substituting payoff values for the hawk-dove game, we have $\sigma(u_{1}) = 2(\mathcal{B}/\mathcal{C})$. For large-$n$, as can be seen from Fig. \ref{plot-sigma}, the phase boundary seems to be reaching an asymptote. We can confer from this observation that for large complex sizes we still expect dove to be the dominant strategy.

\section{Discussion}

The evolution of multicellularity is often treated in a framework that distinguishes (`decouples') the fitness of an individual within a group and the overall fitness of the group, with these quantities deriving from a simple cooperative dilemma such as a linear public goods game.  While this conceptual framework has been very useful in understanding many aspects of the transition to multicellularity \citep{rainey2010cheats}, it can be restrictive in many cases. In particular, it often does not adequately account for the \emph{interactions} of cells within a group, which are likely to be game-theoretic (i.e., frequency-dependent). One particularly clear example is when some cell types are more likely than others to act as reproductive propagules. In the demonstration of \citet{ratcliff2012experimental} that some cells in multicellular yeast clusters undergo apoptosis (ostensibly to help break the group up into several smaller multi-cellular propagules), apoptosed cells obviously cannot propagate further \citep{libby2014ratcheting}. If the likelihood of a cell's apoptosing depends on its genotype, then a model that is not game-theoretic cannot account for this important phenomenon. 

Another example of explicitly game-theoretic interactions within multicellular groups is the competition between `cooperator' and `cheater' strains of yeast which use different glucose metabolism \citep{maclean2006resource, pfeiffer2001cooperation}. In such setups, selection dynamics between different phenotypes with different rates of ATP yield follow a prisoner's dilemma \citep{maclean2006resource}. If staying together (and the concomitant ability to form groups) has already evolved in the population, a model that accounts for this frequency dependence is required if we are adequately to model the evolutionary dynamics of the population.

For these reasons, the construction of a more general, frequency-dependent framework is desirable. In the model developed in this paper, cells divide and stay together until they reach a complex of a certain size. Subsequent cell divisions lead to single cells that leave and start their own complexes. The reproductive rate of a cell is determined by a game based on interactions of cells within a complex, and is therefore explicitly frequency-dependent. This game can represent competition for resources, exchange of nutrients, cellular communication, or energy sharing mechanisms.

Studying the evolutionary dynamics of this setup, we determined how this population structure affects the outcome of evolutionary games in the presence of mutation. In particular, we showed that the condition for one strategy to be more abundant than the other strategy in the mutation-selection equilibrium has the same symmetry as the $\sigma$ condition of stochastic evolutionary dynamics \citep{tarnita2009strategy}. We calculated the value of $\sigma$ and examined it for the game of cooperation and the hawk-dove game.

An important feature of our model is the suppression of the cheater phenotype during the evolution of multicellularity. Since the selection dynamics inside each complex is frequency-dependent (for example, a prisoner's dilemma), a cheater phenotype that appears inside a cooperating colony has a lower chance of becoming abundant inside the group. In other words, a game-theoretic interactions at all levels, individuals and groups, can lead to suppression of the cheater strategy without the need for evolution of new mechanisms of conflict mediation. This is the case when the cooperative phenotype is evolutionarily stable for small mutation rates. Our model also allows us to characterize the critical level of group diversity, the mutation rate $u_{c,1}$, above which a cheater phenotype destabilizes cooperation inside the group.

We see our study primarily as a contribution toward understanding how population structure affects evolutionary dynamics \citep{nowak2010}. We have quantified to what extent the population structure of simple multicellularity, perhaps as found at the various origins of multicellularity, is conducive to favoring cooperation. Cooperation is thought to be crucially involved in evolutionary transitions such as the emergence of multicellularity \citep{maynardsmith1997major,nowak2006evolutionary,nowak2011supercooperators}. Conversely, the somatic evolution of cancer is seen as a breakdown of cooperation among the cells of an organism.

We have allowed throughout for the possibility of very high mutation rates. This is no mere theoretical fancy. While mutations by nucleotide substitution and gene conversion are relatively rare, there are many other sources of frequent mutation.

Mutation in our model is perfectly consistent, for example, with epigenetically-induced heritable phenotype switching, such as that observed in experimental populations of \textit{Pseudomonas} bacteria \citep{beaumont2009experimental}, a model organism in the field of experimental multicellularity \citep{rainey2003evolution, nikolaev2007biofilm, hammerschmidt2014life}. Epigenetic mutations in general, so long as they are heritable, are consistent with our model, and are known in some cases to occur far more frequently than sequence mutations \citep{vandergraaf2015rate}. 

Another source of frequent mutation involves unstable genetic architectures. In \textit{Pseudomonas}, for example,  modularity of the genetic architecture underlying the group-forming phenotype allows it to arise often and in multiple different ways \citep{mcdonald2009adaptive,rainey2010cheats}. 

The directed gene transposition and epigenetic control underlying mating-type switching in yeast \citep{klar1979activation, klar1987differentiated, klar2007lessons}, another model organism in experimental multicellularity \citep{koschwanez2011sucrose, ratcliff2012experimental}, are also consistent with our model. In fission yeast (\textit{Schizosaccharomyces pombe}), mating-type switching is more regular than mutation is in our model \citep{miyata1981mode, egel1984pedigree}, but its high frequency and heritability \citep{klar1987differentiated, klar2007lessons} suggest that genetic mechanisms of the sort underlying it could justify the consideration of very high mutation rates in our model. 

These examples all suggest that high mutation rates might be sufficiently common in organisms undergoing the transition to multicellularity to justify the importance of high mutation rates in our model and its results. It is important to note that mutations in our model must be heritable; for example, non-heritable phenotype switching and cell differentiation in response to environmental cues are ruled out.

Extensions of our model to include cases where staying together is stochastic, which means that cells can leave a given complex with a certain probability, more complicated life cycles, as well as asymmetric mutations are subjects of future works. While our current model is deterministic, the extension to stochastic dynamics and finite total population size should be straightforward.

Our model does not study important questions such as the different implications of staying together versus coming together \citep{tarnita2013} or the evolution of germ line soma separation \citep{michod2003reorganization,michod2001cooperation} or the evolution of simple versus complex multicellularity \citep{knoll2011multiple} for which we refer to the existing literature.

\section{Acknowledgments}
Support from the John Templeton Foundation is gratefully acknowledged.

\appendix
\section{Exact solutions for $n=2$}
In this appendix we present derivation of equilibrium solutions for $n=2$. Putting time
derivatives to zero in Eq. \ref{gm-mult}, we have

\bea
x^{\star}_{0} &=& \frac{1}{\phi^{\star}+1}\big(P_{00}x^{\star}_{00}+P_{01}x^{\star}_{01}+P_{11}x^{\star}_{11}\big)\nonumber\\
x^{\star}_{1} &=& \frac{1}{\phi^{\star}+1}\big(Q_{00}x^{\star}_{00}+Q_{01}x^{\star}_{01}+Q_{11}x^{\star}_{11}\big)\nonumber\\
x^{\star}_{00} &=& \frac{(1-u)x^{\star}_{0}}{\phi^{\star}}\nonumber\\
x^{\star}_{01} &=& \frac{u(x^{\star}_{0}+x^{\star}_{1})}{\phi^{\star}}\nonumber\\
x^{\star}_{11} &=& \frac{(1-u)x^{\star}_{1}}{\phi^{\star}}
\label{gm-mult-app}
\eea

\nd Substituting for complex abundances from last three equations into first two equations in \ref{gm-mult-app} and dividing them we obtain the following relation for $\eta^{\star} = x^{\star}_{1}/x^{\star}_{0}$

\be
\eta^{\star}= \displaystyle\frac{\big(Q_{00}(1-u)+Q_{01}u\big) + \big(Q_{11}(1-u)+Q_{01}u\big)\eta^{\star}}{\big(P_{00}(1-u)+P_{01}u\big) + \big(P_{11}(1-u)+P_{01}u\big)\eta^{\star}} 
\label{eta-star}
\ee

\nd Average fitness function at equilibrium, $\phi^{\star}$, can be expressed in terms of $\eta^{\star}$ as well

\bea
(\phi^{\star})^{2} &+& \phi^{\star} =  \frac{1}{1 + \eta^{\star}}\Big\{\big(P_{00}+Q_{00}\big)(1-u)+\big(P_{01}+Q_{01}\big)u \nonumber\\
 &+&\Big(\big((P_{11}+Q_{11}\big)(1-u)+\big(P_{01}+Q_{01}\big)u\Big)\eta^{\star} \Big\}
\label{phi-star}
\eea

\nd The above equations, \ref{eta-star} and \ref{phi-star}, can be rewritten as quadratic equations

\bea
A(\eta^{\star})^{2} + B\eta^{\star} - C = 0\nonumber\\
(\phi^{\star})^{2} + \phi^{\star} - \frac{1}{1 + \eta^{\star}}\Big(D_{0}+D_{1}\eta^{\star}\Big)=0
\label{quad-eta-phi}
\eea

\nd Coefficients $A,B,C,D_{0}$ and $D_{1}$ are expressed in terms of payoff coefficients and mutation rate

\bea
A &\equiv& (1-u)P_{11}+uP_{01}\nonumber\\
&=& \Big(\big( -2+w(c-b-2d)\big)u +\big((b+2d)w+3\big)\Big)u\nonumber\\
B &\equiv& (P_{00}-Q_{11})(1-u)+(P_{01}-Q_{01})u\nonumber\\
&=& \big( 2w(a-b+c-d\big)u^{2}+w\big(4(d-a)+b-c)u+2w(a-d)\nonumber\\
C &\equiv& Q_{00}(1-u)+Q_{01}u\nonumber\\
&=& \big((b-2a-c)w -2\big)u^{2}+\big((2a+c))w+3\big)u \nonumber\\
D_{0} &\equiv& \Big((P_{00}+Q_{00})(1-u) + (P_{01}+Q_{01})u\Big) \nonumber\\
&=& 2(1+wa)(1-u) + \big(2+w(b+c)\big)u\nonumber\\
D_{1} &\equiv&  \Big((P_{11}+Q_{11})(1-u) + (P_{01}+Q_{01})u\Big) \nonumber\\
&=& 2(1+wd)(1-u) + \big(2+w(b+c)\big)u\nonumber\\ 
\label{ABC}
\eea

\nd Values of $\eta^{\star}$ and $\phi^{\star}$ are thus given by

\bea
\eta^{\star}&=&\frac{-B+\sqrt{B^{2}+4AC}}{2A}\nonumber\\
\phi^{\star}&=&\frac{-1+\sqrt{\displaystyle 1+4\frac{(D_{0} + D_{1}\eta^{\star})}{1 + \eta^{\star}}}}{2}
\label{phi-eta1}
\eea

Coefficients $A,C,D_{0}$ and $D_{1}$ in Eq. \ref{ABC} are positive since coefficients $P_{00},P_{01}$, $P_{11}$
and  $Q_{00}, Q_{01}, Q_{11}$ are positive for $0 < u < 1$. Both $\eta^{\star}$ as the ratio of two abundance, and  $\phi^{\star}$ as average fitness should be positive as well. Thus solutions for Eq. \ref{phi-eta1} are unique positive solutions of Eq. \ref{quad-eta-phi}. Given $\eta^{\star}$ and $\phi^{\star}$, abundances of singlet and doublet complexes at equilibrium $x^{\star}_{0},x^{\star}_{1},x^{\star}_{00},x^{\star}_{01},x^{\star}_{11}$ are

\bea
x^{\star}_{0} &=& \frac{\phi^{\star}}{2 + \phi^{\star}}\frac{1}{1+\eta^{\star}}\nonumber\\
x^{\star}_{1}&=& \frac{\phi^{\star}}{2 + \phi^{\star}}\frac{\eta^{\star}}{\eta^{\star} + 1}\nonumber\\
x^{\star}_{00}&=&\frac{1-u}{2+\phi^{\star}}\frac{1}{1+\eta^{\star}}\nonumber\\
x^{\star}_{01}&=&\frac{u}{2 +\phi^{\star}}\nonumber\\
x^{\star}_{11}&=&\frac{1-u}{2+\phi^{\star}}\frac{\eta^{\star}}{1+\eta^{\star}}\nonumber\\
\label{equil-app}
\eea

\nd We used $x^{\star}_{0}+x^{\star}_{1}= \phi^{\star}/(2 + \phi^{\star})$. This can be checked by substituitng 
steady state solutions $x^{\star}_{00},x^{\star}_{01},x^{\star}_{11}$ from Eq. \ref{gm-mult-app} into identity $x^{\star}_{0}+x^{\star}+2(x^{\star}_{00}+x^{\star}_{01}+x^{\star}_{11})=1$.

\section{Evolutionary stability for $n=2$}
In this appendix we present some technical details of the results on stability 
of the equilibrium solutions and ESS condition for $n=2$. We write Eq. \ref{gm-mult} as

\be
\frac{{\rm d}{x}_{i}}{{\rm d}t} = F_{i}\big(x_{0},x_{1},x_{00},x_{11}\big)
\label{b1}
\ee

\nd where $i \in \{ 0, 1, 00, 11\}$. Mixed complex $x_{01}$ is expressed in terms of other
variables

\be
x_{01} = \frac{1}{2}\Big( 1 - x_{0}-x_{1}-2x_{00}-2x_{11}\Big)	
\ee

To address stability we linearize time derivative operator, $F_{i}\big(x_{0},x_{1},x_{00},x_{11}\big)$ around a fixed point. 
For a stable fixed point, all real parts of eigenvalues of the linearized $F_{i}\big(x_{0},x_{1},x_{00},x_{11}\big)$, i.e. Jacboian, should be negative. At $u=0$ and for zero-selection, $w=0$, average fitness $\phi=1$. Jacobian has following eigenvalues and eigenvectors: $\lambda^{(0)}_{1}=0, \vec{v}_{1}=(-1,1,-1,1), \lambda^{(0)}_{2}=-1, \vec{v}_{2}=(0,0,1,1), \lambda^{(0)}_{3}=-3, \vec{v}_{3}=(2,-2,-1,1), \lambda^{(0)}_{4}=-3, \vec{v}_{4}=(2,2,-1,-1)$. At weak selection the above eigen-directions are slightly modified due to game payoff contribution to the fitness of type 0 and type 1 strategies. Particularly, the direction corresponding to $\lambda^{(0)}_{1}=0$ can become unstable. 

To check this we focus on the type 0 fixed point $(x^{\star}_{0},0,x^{\star}_{00},0)$ (denoted by $(i)$). We 
perturb it along the $\vec{v}_{1}$ direction 

\bea
x_{0} =x^{\star}_{0}&\longrightarrow& x^{\star}_{0} - \frac{\delta x}{3}\nonumber\\
x_{1}=0&\longrightarrow &\frac{\delta x}{3}\nonumber\\
x_{00} =x^{\star}_{00}&\longrightarrow& x^{\star}_{00}-\frac{\delta x}{3} \nonumber\\
x_{01}=0&\longrightarrow& 0\nonumber\\
x_{11}=0&\longrightarrow& \frac{\delta x}{3}\nonumber\\
\eea

\nd This corresponds to introducing a small fraction $\delta x$  of type 1 cells to the system: $x_{\rm tot,1} \to \delta x$, $x_{\rm tot,0} \to 1-\delta x$. Substituting into Eq. \ref{gm-mult} (or Eq. \ref{b1}) and keeping terms up to lowest order in $w$ we obtain linearized equation
around the fixed point $(x^{\star}_{0},x^{\star}_{1}=0,x^{\star}_{00},x^{\star}_{01}=0,x^{\star}_{11}=0)$

\be
\frac{{\rm d}\delta x}{{\rm d}t} \approx -\frac{2w}{3}\big(a-d\big)\delta x + \mathcal{O}(w^{2})
\ee

\nd which indicates corresponding eigenvalue $\lambda^{(i)}_{1}\approx -(2w/3)(a-d)$. For $a > d$ and independent of off-diagonal payoff coefficients, $b$ and $c$, a fixed point of $(x^{\star}_{0},0,x^{\star}_{00},0)$ is ESS. Similar condition can be obtained by linearizing time 
opertaor around type 1 fixed point, $(0,x^{\star}_{1},0,x^{\star}_{11})$. This leads to eigenvalue, $\lambda^{(ii)}_{1}\approx(2w/3)(a-d)$ ($(ii)$ denotes type 1 fixed point). Thus for the Jacobian matrix, $J_{ij}$, defined as

\be
J_{ij} = \frac{\partial F_{i}}{\partial x_{j}}\at{\vec{x}=\vec{x}^{\star}},~~~~~\vec{x}=(x_{0},x_{1},x_{00},x_{11})
\ee

\nd all the eigenvalues can be calculated to the leading order of selection intensity. For type 0 fixed point we obtain

\bea
\lambda^{(i)}_{1}&\approx& -\frac{2}{3}\big( a-d\big)w + \mathcal{O}\big(w^{2}\big)\nonumber\\
\lambda^{(i)}_{2}&\approx& -1-\frac{2}{3}aw+ \mathcal{O}\big(w^{2}\big)\nonumber\\
\lambda^{(i)}_{3}&\approx&  -3 + \Big( a - \frac{1}{3}d+|a-d|\Big)w + \mathcal{O}\big(w^{2}\big)\nonumber\\
\lambda^{(i)}_{4}&\approx& -3 + \Big( a - \frac{1}{3}d -|a-d|\Big)w + \mathcal{O}\big(w^{2}\big)\nonumber\\
\label{eigen-weak}
\eea

\begin{figure}
\begin{center}
\epsfig{figure=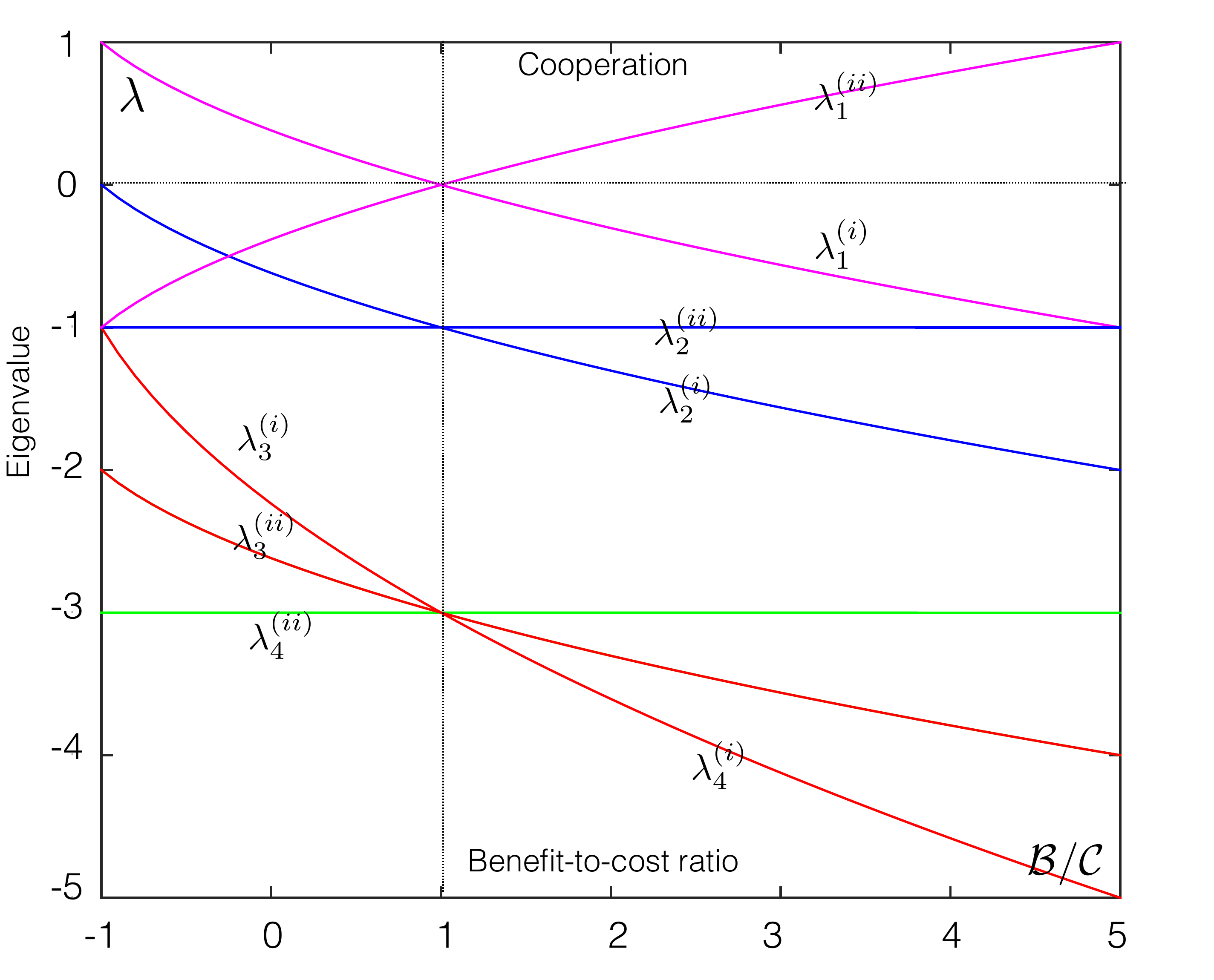, height=220pt,width=250pt,angle=0}
\epsfig{figure=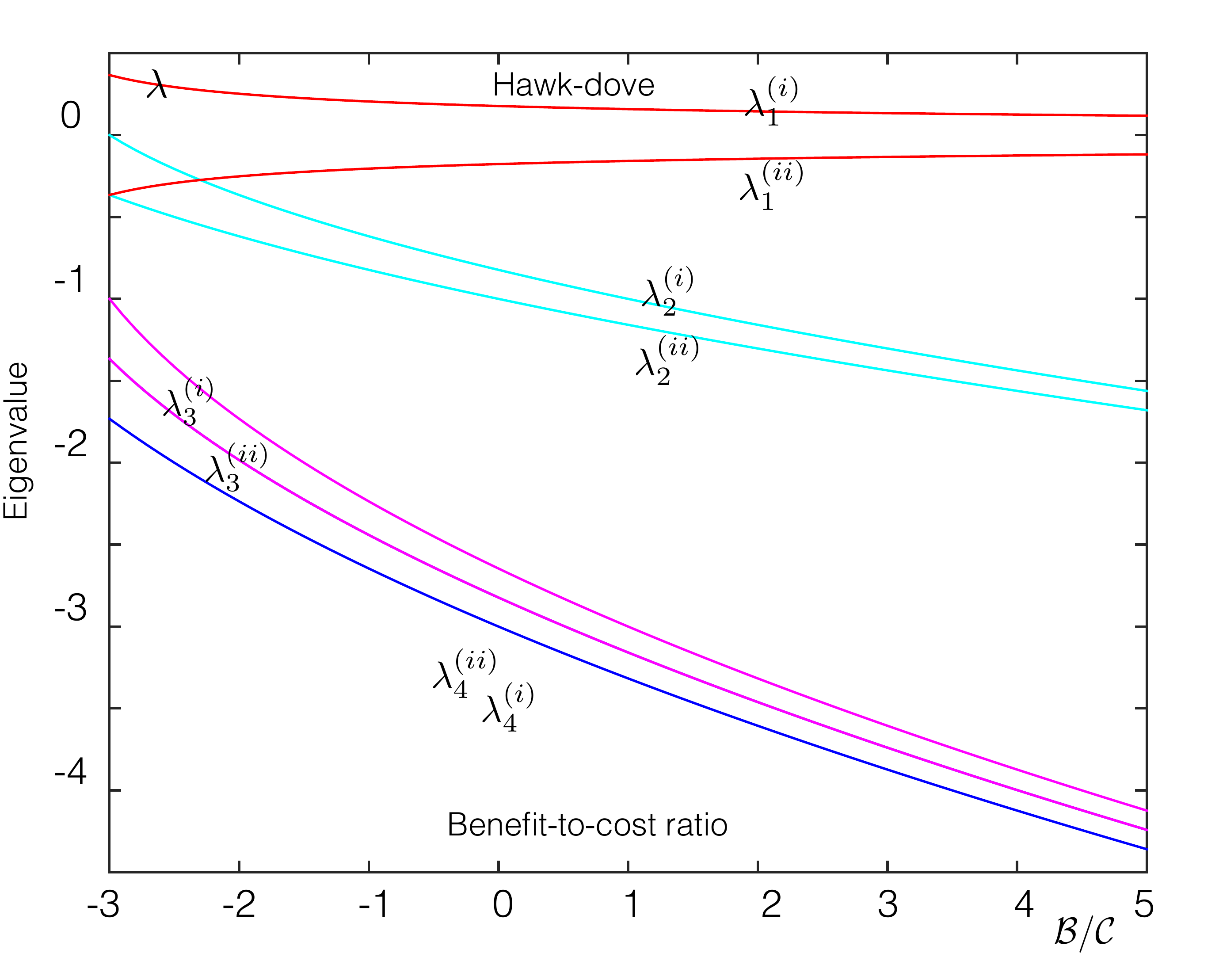, height=220pt,width=250pt,angle=0}
\end{center}
\caption{Numerical results for eigenvalues for the game of cooperation (top) and haw-dove game (bottom). The eigenvalues are plotted as a function of $\mathcal{B}/\mathcal{C}$ for $u=0$ and $w=0.5$. 
 Eigenvalues of type 0 fixed point, 
$\lambda^{(i)}_{1,2,3,4}$, and type 1 fixed point, $\lambda^{(ii)}_{1,2,3,4}$, are plotted. ESS condition for type 0 is $\mathcal{B}/\mathcal{C} > 1$ for the game of cooperation. As can be seen at value $\mathcal{B} = \mathcal{C}$, $\lambda^{(i.ii)}_{1}$ changes sign. For the hawk-dove game however, dove strategy is always ESS at $u=0$.}
\label{eigen-hawk-coop}
\end{figure}

We numerically calculated all the eigenvalues around both fixed-points at $u=0$. For $w=0.01$ results matched very well with Eq. \ref{eigen-weak}. For larger intensities of selection there are deviations from Eq. \ref{eigen-weak}. However it seems that the ESS condition, $a > d$, still holds away from weak selection as well. Fig. \ref{eigen-hawk-coop} shows numerical results for eigenvalues around both fixed points at $u=0$ for the game of cooperation and the hawk-dove game. Eigenvalues are plotted as a function of $\mathcal{B}/\mathcal{C}$ and for $w=0.5$. For the game of cooperation as the benefit to cost ratio passes unity, $\mathcal{B}/\mathcal{C}=1$, the eigenvalue $\lambda^{(i)}_{1}$ switches sign thus indicting the type 0 is evolutionary stable for $\mathcal{B} > \mathcal{C}$. 

For $n>2$ the above analysis can be tedious but in principle the same steps can be done. The same result, however, can be intuitively understood. For finite $u$ there is a single equilibrium state that is globally attractive inside the multi-dimensional simplex of states. As $u \to 0$ 
this fixed point moves approaches to the ESS fixed point among the two fixed point for $u=0$. This is indicated by the $\sigma$-condition. The
strategy that is favoured by $\sigma$-condition as $u \to 0$ is the ESS. 

\section{Exact solutions for $n=3$}
Here we present a sketch of derivation of exact solutions for model with maximum three-cell complexes. 
The equilibrium abundances for $n=3$ satisfy the coupled system of equation:

\bea
x^{\star}_{1,0}&=& \frac{1}{1 + \phi^{\star}}\Big( P_{3,0}x^{\star}_{3,0}+P_{2,1}x^{\star}_{2,1}+P_{1,2}x^{\star}_{1,2}+P_{0,3}x^{\star}_{0,3}\Big)\nonumber\\
x^{\star}_{0,1}&=&\frac{1}{1 + \phi^{\star}}\Big( Q_{3,0}x^{\star}_{3,0}+Q_{2,1}x^{\star}_{2,1}+Q_{1,2}x^{\star}_{1,2}+Q_{0,3}x^{\star}_{0,3}\Big)\nonumber\\
x^{\star}_{2,0}&=&\frac{1}{P_{2,0}+Q_{2,0}+\phi^{\star}}P_{1,0}x^{\star}_{1,0} \nonumber\\
x^{\star}_{1,1}&=&\frac{1}{P_{1,1}+Q_{1,1}+\phi^{\star}}\Big( P_{0,1}x^{\star}_{0,1}+Q_{1,0}x^{\star}_{1,0}\Big)\nonumber\\
x^{\star}_{0,2}&=&\frac{1}{P_{0,2}+Q_{0,2}+\phi^{\star}}Q_{0,1}x^{\star}_{0,1}\nonumber\\
x^{\star}_{3,0}&=&\frac{1}{\phi^{\star}}P_{2,0}x^{\star}_{2,0}\nonumber\\
x^{\star}_{2,1}&=&\frac{1}{\phi^{\star}}\Big( P_{1,1}x^{\star}_{1,1}+Q_{2,0}x^{\star}_{2,0}\Big)\nonumber\\
x^{\star}_{1,2}&=&\frac{1}{\phi^{\star}}\Big( P_{0,2}x^{\star}_{0,2}+Q_{1,1}x^{\star}_{1,1}\Big)\nonumber\\
x^{\star}_{0,3}&=&\frac{1}{\phi^{\star}}Q_{0,2}x^{\star}_{0,2}
\label{gm-mult-n3}
\eea

\nd Coefficients $P_{i,j}$ and $Q_{i,j}$ are given by  Eq. \ref{P-gen}. We also have $P_{1,0}=Q_{0,1}=1-u, P_{0,1}=Q_{1,0}=u$. Similar to $n=2$ case, solutions $x^{\star}_{i,3-i}$ can be 
expressed in terms of ratio of singlets $\eta^{\star}=x^{\star}_{1}/x^{\star}_{0}$ and total fitness $\phi^{\star}$. Upon dividing the first two equations in Eq. \ref{gm-mult-n3} and substituting for abundances from rest of the equations we obtain a quadratic equation for $\eta^{\star}$

\bea
\eta^{\star} &=& \Bigg\{ Q_{3,0}\Big(\frac{P_{2,0}}{\phi^{\star}+P_{2,0}+Q_{2,0}}(1-u)\Big) + Q_{2,1}\Big( P_{1,1}\frac{u}{\phi^{\star}+P_{1,1}+Q_{1,1}}(1+\eta^{\star})\nonumber\\ 
&+& \frac{Q_{2,0}(1-u)}{\phi^{\star}+P_{2,0}+Q_{2,0}}\Big) + Q_{1,2}\Big( \frac{P_{0,2}(1-u)}{\phi^{\star}+P_{0,2}+Q_{0,2}}\eta^{\star} + \frac{Q_{1,1}u}{\phi^{\star}+P_{1,1}+Q_{1,1}}(\eta^{\star}+1)\Big)\nonumber\\
&+& Q_{0,3}\Big( \frac{Q_{0,2} (1-u)}{\phi^{\star}+P_{0,2}+Q_{0,2}}\eta^{\star}\Big)\Bigg\} \Bigg / 
 \Bigg\{P_{3,0}\Big(\frac{P_{2,0}}{\phi^{\star}
 +P_{2,0}+Q_{2,0}}(1-u)\Big) \nonumber\\
 &+&P_{2,1}\Big( P_{1,1}\frac{u}{\phi^{\star}+P_{1,1}+Q_{1,1}}(1+\eta^{\star})
+ \frac{Q_{2,0}(1-u)}{\phi^{\star}+P_{2,0}+Q_{2,0}}\Big)\nonumber\\
 &+& P_{1,2}\Big( \frac{P_{0,2}(1-u)}{\phi^{\star}+P_{0,2}+Q_{0,2}}\eta^{\star} + \frac{Q_{1,1}u}{\phi^{\star}+P_{1,1}+Q_{1,1}}(\eta^{\star}+1)\Big)\nonumber\\
&+& P_{0,3}\Big( \frac{Q_{0,2} (1-u)}{\phi^{\star}+P_{0,2}+Q_{0,2}}\eta^{\star}\Big)\Bigg\}
\label{eta-n3}
\eea

\nd singlet and complex abundances, up to the common factor, $x^{\star}_{0,1}+x^{\star}_{1,0}$, are expressed in terms of $\eta^{\star}$ and $\phi^{\star}$ from Eq. \ref{gm-mult-n3},

\bea
x^{\star}_{2,0}&=&\frac{1-u}{\phi^{\star}+P_{2,0}+Q_{2,0}}\frac{1}{1 + \eta^{\star}}\big(x^{\star}_{1,0}+x^{\star}_{0,1}\big)\nonumber\\
x^{\star}_{1,1}&=&\frac{u}{\phi^{\star}+P_{1,1}+Q_{1,1}}\big(x^{\star}_{1,0}+x^{\star}_{0,1}\big)\nonumber\\
x^{\star}_{0,2}&=&\frac{1-u}{\phi^{\star}+P_{0,2}+Q_{0,2}}\frac{\eta^{\star}}{1 + \eta^{\star}}\big(x^{\star}_{1,0}+x^{\star}_{0,1}\big)\nonumber\\
x^{\star}_{3,0}&=&\frac{1}{\phi^{\star}}P_{2,0}\Big(\frac{1-u}{\phi^{\star}+P_{2,0}+Q_{2,0}}\frac{1}{1 + \eta^{\star}}\Big)\big(x^{\star}_{1,0}+x^{\star}_{0,1}\big)\nonumber\\
x^{\star}_{2,1}&=&\frac{1}{\phi^{\star}}\Big(P_{1,1}\frac{u}{\phi^{\star}+P_{1,1}+Q_{1,1}}+ Q_{2,0}\frac{1-u}{\phi^{\star}+P_{2,0}+Q_{2,0}}\frac{\eta^{\star}}{1 + \eta^{\star}}\Big)\big(x^{\star}_{1,0}+x^{\star}_{0,1}\big)\nonumber\\
x^{\star}_{1,2}&=&\frac{1}{\phi^{\star}}\Big( P_{0,2}\frac{1-u}{\phi^{\star}+P_{0,2}+Q_{0,2}}\frac{\eta^{\star}}{1 + \eta^{\star}}+
Q_{1,1}\frac{u}{\phi^{\star}+P_{1,1}+Q_{1,1}}\Big)\big(x^{\star}_{1,0}+x^{\star}_{0,1}\big)\nonumber\\
x^{\star}_{0,3}&=&\frac{1}{\phi^{\star}}Q_{0,2}\frac{1-u}{\phi^{\star}+P_{0,2}+Q_{0,2}}\frac{\eta^{\star}}{1 + \eta^{\star}}\big(x^{\star}_{1,0}+x^{\star}_{0,1}\big)\nonumber\\
\label{gm-mult-sol-n3}
\eea

\nd Substituting these into condition  $x^{\star}_{1,0}+x^{\star}_{0,1}+2\big(x^{\star}_{2,0}+x^{\star}_{1,1}+x^{\star}_{0,2}\big)+3(x^{\star}_{3,0}+x^{\star}_{2,1}+x^{\star}_{1,2}+x^{\star}_{0,3}\big)=1$, we obtain $x^{\star}_{1,0} + x^{\star}_{0,1}$ in terms of $\eta^{\star}$ and $\phi^{\star}$

\bea
x^{\star}_{1,0}+x^{\star}_{0,1} &=& \phi^{\star}\Bigg/ \Bigg\{\phi^{\star} + \frac{1}{1 + \eta^{\star}}\times
\Bigg[\Big( \big(3(P_{2,0}+Q_{2,0}\big)+2\phi^{\star}\Big)\frac{1-u}{\phi^{\star}+P_{2,0}+Q_{2,0}}\nonumber\\
&&+\Big( \big(3(P_{1,1}+Q_{1,1}\big)+2\phi^{\star}\Big)\frac{u(1+\eta^{\star})}{\phi^{\star}+P_{1,1}+Q_{1,1}}\nonumber\\
&&+\Big( \big(3(P_{0,2}+Q_{0,2}\big)+2\phi^{\star}\Big)\frac{(1-u)\eta^{\star}}{\phi^{\star}+P_{0,2}+Q_{0,2}}\Big)\Bigg]\Bigg\}\nonumber\\
\label{x0+x1}
\eea

$\phi^{\star}$ is obtained from substituting above solutions, Eq. \ref{gm-mult-sol-n3} and Eq. \ref{x0+x1} into
\bea
\phi^{\star} &=& \big( P_{3,0}+Q_{3,0}\big) x^{\star}_{3,0}+\big( P_{2,1}+Q_{2,1}\big) x^{\star}_{2,1}\nonumber\\
&+&\big( P_{1,2}+Q_{1,2}\big) x^{\star}_{1,2}+\big( P_{0,3}+Q_{0,3}\big) x^{\star}_{0,3}\nonumber\\
&+&\big( P_{2,0}+Q_{2,0}\big) x^{\star}_{2,0}+\big( P_{1,1}+Q_{1,1}\big) x^{\star}_{1,1}\nonumber\\
&+&\big( P_{0,2}+Q_{0,2}\big) x^{\star}_{0,2}+x^{\star}_{1,0}+x^{\star}_{0,1}\nonumber\\
\label{phi-n3}
\eea

Eq. \ref{eta-n3} and Eq. \ref{phi-n3} combined with Eq. \ref{x0+x1} can be solved to obtain closed form solutions for $\phi^{\star}$ and $\eta^{\star}$, which upon subtituting into Eq. \ref{gm-mult-sol-n3} gives all abundances in terms of payoff values and mutation. This approach in principle is generalized to $n$-cell complex model as well. We always get a quadratic equation in $\eta^{\star}$ coupled with
a degree-$n$ equation for $\phi^{\star}$. The condition for the strategy selection, however, can be answered in 
weak selection limit without exact knowledge of individual abundance of different complexes as discussed in Sec.5.

\section{Well-mixed games with mutation}
In this appendix we briefly review unstructured game in the presence of mutations and compare the condition for neutrality with the one 
we obtained for game of multicellularity, Eq. \ref{gm-mult}. Derivations for finite populations are done in literature
\citep{tarnita2009strategy, traulsen2008analytical, antal2009strategy}. Consider a well-mixed population of two populations 0 and 1 with abundances $x_{0}$ and $x_{1}$. Each individual can replicate based on its payoff values determined by the game and the offspring can mutate with probability $u$. The dynamics for two populations is written as,

\bea
\dot{x}_{0}&=& x_{0}\Big(P_{0}x_{0}+P_{1}x_{1}\Big)(1-u) + x_{1}\Big(Q_{0}x_{0}+Q_{1}x_{1}\Big)u - x_{0}-x_{0}\phi\nonumber\\
\dot{x}_{1}&=& x_{0}\Big(P_{0}x_{0}+P_{1}x_{1}\Big)u + x_{1}\Big(Q_{0}x_{0}+Q_{1}x_{1}\Big)(1-u)-x_{1}-x_{1}\phi\nonumber\\
\eea

\nd where fitness functions, $P_{0,1}, Q_{0,1}$, are given in terms of payoff coefficients $a,b,c,d$, and intensity of selection $w$

\bea
P_{0}&=& (1+wa)\nonumber\\ 
P_{1}&=& (1+wb)\nonumber\\
Q_{0}&=& (1+wc)\nonumber\\
Q_{1}&=& (1+wd)
\eea

\nd $\phi$ is given by

\be
\phi = x_{0}\big( P_{0}x_{0}+P_{1}x_{1}\big) + x_{1}\big( Q_{0}x_{0}+Q_{1}x_{1}\big) -1
\label{phi-un}
\ee

\nd Enforcing the condition that total frequencies add up to unity, $x_{0}+x_{1}=1$. Calling $x_{0}=x(t)$, we have

\be
\frac{\rm d}{{\rm d}t}x(t) = x\big( 1 - u -x\big)\Big( P_{0}x + P_{1}(1-x)\Big) + (1-x)\big(u-x\big)\Big(Q_{0}x + Q_{1}(1-x)\Big),
\ee

\nd Putting LHS of the above equation to zero for equilibrium solutions $(x^{\star}$) and  assuming type 0 is selected, i.e., $x^{\star}> 1/2$, after some straightforward algebra we get

\be
a + b > c + d
\label{uc-well-mixed}
\ee

\nd which is the $\sigma$-condition with $\sigma=1$. Notice that this relationship is independent of $u$.

\section*{References}
\bibliographystyle{spbasic}      
\bibliography{mybib-game-multicell.bib}   

\end{document}